\documentclass[fleqn,usenatbib]{mnras}

\usepackage{newtxtext,newtxmath}

\usepackage[T1]{fontenc}
\usepackage{ae,aecompl}

\usepackage{caption}
\usepackage{subcaption}
\usepackage{float}
\usepackage{graphicx}	
\usepackage{amsmath}	
\usepackage{amssymb}	

\usepackage{hyperref}
\usepackage{xcolor}
\usepackage{ulem}

\newcommand{\orcid}[1]{\href{https://orcid.org/#1}{\textcolor[HTML]{A6CE39}{\aiOrcid}}}

\title[Mock FRB injection at UTMOST]{Estimating fast transient detection pipeline efficiencies at UTMOST via real-time injection of mock FRBs}

\author[Gupta, V. et al.]{\parbox{\textwidth}
   {V.\ Gupta$^{1}$, 
    C.\ Flynn$^{1,2}$,
    W.\ Farah$^{1, 8}$,
    A.\ Jameson$^{1, 2}$, 
    V.\ Venkatraman Krishnan$^{1,6}$, 
    M.\ Bailes$^{1,2}$,
    T.\ Bateman$^{1, 3}$, 
    A.\ T.\ Deller$^{1, 2}$,
    A.\ Mandlik$^{1}$,
    A.\ Sutherland$^{1, 3}$
    }  \\ \\ \\
\parbox{\textwidth}{
$^{1}$Centre for Astrophysics and Supercomputing,
  Swinburne University of Technology, Mail H30, PO Box 218, VIC 3122,
  Australia\\
$^{2}$ARC Centre of Excellence for Gravitational Wave Discovery (OzGrav), Australia\\
$^{3}$Sydney Institute for Astronomy, School of Physics A28, University
  of Sydney, NSW 2006, Australia\\
$^{4}$CSIRO Astronomy and Space Science, Australia Telescope National Facility, Epping, NSW 1710, Australia\\
$^{5}$Center for Astrophysics | Harvard \& Smithsonian, 60 Garden Street, Cambridge MA 02138, USA\\
$^{6}$Max-Planck-Institute f\"{u}r Radioastronomie, Auf dem H\"{u}gel 69, D-53121 Bonn, Germany\\
$^{8}$SETI Institute 189 Bernardo Ave, Suite 200 Mountain View, CA 94043, United States\\
}
}

\date{Accepted XXX. Received YYY; in original form ZZZ}

\pubyear{2020}

\begin{document}
\label{firstpage}
\pagerange{\pageref{firstpage}--\pageref{lastpage}}
\maketitle

\newcommand{\dm}{$\textrm{pc}\,\textrm{cm}^{-3}$}

\begin{abstract}
Dedicated surveys using different detection pipelines are being carried out at multiple observatories to find more Fast Radio Bursts (FRBs). Understanding the efficiency of detection algorithms and the survey completeness function is important to enable unbiased estimation of the underlying FRB population properties.
One method to achieve end-to-end testing of the system is by injecting mock FRBs in the live data-stream and searching for them blindly. Mock FRB injection is particularly effective for machine-learning-based classifiers, for which analytic characterisation is impractical. We describe a first-of-its-kind implementation of a real-time mock FRB injection system at the upgraded Molonglo Observatory Synthesis Telescope (UTMOST) and present our results for a set of 20,000 mock FRB injections. 
The injections have yielded clear insight into the detection efficiencies and have provided a survey completeness function for pulse width, fluence and DM.
Mock FRBs are recovered with uniform efficiency over the full range of injected DMs, however the recovery fraction is found to be a strong function of the width and Signal-to-Noise (SNR). For low widths ($\lesssim 20$ ms) and high SNR ($\gtrsim$ 9) the recovery is highly effective with recovery fractions exceeding 90\%.
We find that the presence of radio frequency interference causes the recovered SNR values to be systematically lower by up to 20\% compared to the injected values. We find that wider FRBs become increasingly hard to recover for the machine-learning-based classifier employed at UTMOST.
We encourage other observatories to implement live injection set-ups for similar testing of their surveys.
\end{abstract}

\begin{keywords}
(transients:) fast radio bursts -- instrumentation: interferometers -- methods: data analysis
\end{keywords}



\section{Introduction}

First discovered in 2007 \citep{Lorimer2007}, Fast Radio Bursts (FRBs) have become a very active field of research in the past decade \citep{Petroff2019_review, Cordes2019_review}. These bright, millisecond-duration bursts are now established to be originating at cosmological distances, subsequently carrying an imprint of the Intergalactic Medium along their propagation path, offering a unique probe into the structure, composition, magnetic field and other properties of the ionised baryons present along the line of sight \citep{Macquart_baryons2020,Macquart2017, Zheng2014, Prochaska2019_haloes}.

A total of 117 unique FRB detections have been published to date (August 2020; \citealp{Petroff2017}), some 20 of which have been observed to produce repeat bursts. Numerous projects are underway to build new telescopes or upgrade existing ones to find and localize more FRBs \citep{DSA-10, CHIME_description, Realfast, greenburst_surnis2019, meertrap_stappers2018}.
Notably, the Australian Square Kilometer Array Pathfinder (ASKAP) telescope, Deep Synoptic Array-110 (DSA-110), Canadian Hydrogen Intensity Mapping Experiment (CHIME) and upgraded Molonglo Observatory Synthesis Telescope (UTMOST-2D) promise a large number of FRB detections in future and are currently under development or in a commissioning phase.

A defining characteristic of FRBs is the frequency dependent delay introduced by wave propagation through a cold, ionised medium. This delay is parameterized by the Dispersion Measure (DM), which is proportional to the integrated column density of electrons along the line of sight and is measured in pc cm$^{-3}$. This dispersion, if not properly accounted and corrected for, will result in the signal getting smeared out below the detection threshold. Since the DM of FRBs is not known \textit{a priori}, finding them requires searching through a large number of DM trials across the desired DM range.
Combined with the high time resolution ($\sim$millisecond) required, this typically makes the parameter space so large that it is impractical for all candidates to be scrutinised manually.

A number of automated FRB search pipelines have been developed over the last decade. At the core of most search pipelines lie dedispersion and thresholding algorithms (e.g. \textsc{SEEK}\footnote{\href{SEEK}{http://sigproc.sourceforge.net/seek/}},  \textsc{HEIMDALL}\footnote{\href{HEIMDALL}{https://sourceforge.net/p/heimdall-astro/wiki/Home/}}, \textsc{bonsai}\footnote{\cite{CHIME_system2018}}, etc.) which remove the dispersion delay for a set of DM trials and produce a list of candidates above a significance threshold.  


Prior to automation, FRBs were found manually by inspecting the list of candidates output by a dedispersion and thresholding algorithm. Three algorithms have been widely used, namely SEEK \citep{Lorimer2007}, HEIMDALL \citep{Champion2015} and PRESTO\footnote{\href{https://www.cv.nrao.edu/~sransom/presto/}{https://www.cv.nrao.edu/~sransom/presto/}} \citep{Spitler2014}.

However, the presence of Radio Frequency Interference (RFI) can lead to the generation of a large number of false positive signals when using basic thresholding search methods. This makes it challenging if a human is faced with classifying thousands of candidates daily.
Ranking the candidates can make the human verification process more efficient and reduce the effort to manageable levels. For example, \cite{Karako-Argaman_RRATrap_sifting} developed a single-pulse sifting algorithm \texttt{RRAT}trap making use of the properties that characterize astrophysical signals to give high ranks to the most promising candidates, which were then selected for human verification.

More recently, machine learning techniques have been employed at facilities searching for FRBs to reduce the number of false positives, training on signals which have an astronomical origin (such as single pulses from pulsars;  \citealp{Farah2018, Agarwal2019_ML, Zhang2018_ML, Michilli2018}). Removing the need for a human classifier also enables these pipelines to work in real-time and add the capability to trigger capture of raw voltages even for telescopes heavily affected by false positives from RFI, thereby allowing coherent dedispersion and high time resolution studies of detected FRBs \citep{Farah2018}.

Understanding the efficiency and completeness of these algorithms is crucial to the reliability of the results derived from the FRB detections. As \cite{Keane-Petroff2015} have demonstrated, the derived fluences, DMs and SNRs of FRBs, as well as the implied detection rates will all be affected if the detection efficiency and parameter recovery efficiency have not been accounted for. 

For example, \cite{Farah2019MNRAS} recently estimated the FRB sky rate at 843 MHz with UTMOST, finding it significantly lower than at 1.4 GHz as scaled from the observed sky rates at ASKAP \citep{Shannon2018} and Parkes \citep{Bhandari2018}. They interpreted the result as hinting at a spectral turnover in the energy distribution of FRBs at frequencies between 843 MHz and 1.4 GHz. In order to establish confidence in the result, they needed to perform a rigorous validation of the detection pipeline.

Efforts have been made by developers and users to characterise the efficiency and completeness of the detection algorithms individually \citep{Keane-Petroff2015, Barsdell_HEIMDALL, CHIME_system2018, Connor2018}, and more recently, end-to-end testing systems are being implemented by various groups, to test performance of the complete pipelines using mock FRB injection.

\cite{Patel2018_PALFA_survey} and \cite{Aggarwal2020_GREENBURST} measured the detection pipeline efficiencies at the Arecibo and GBT telescopes respectively. They run their detection pipelines on archival data-sets injected with mock FRBs to assess the pipeline performance  over a range of SNR, DM and widths and report the end-to-end efficiency of FRB detection from their pipelines. We are aware that mock FRB injection systems have also been developed for CHIME \citep{CHIME_description}, \textit{realfast} \citep{Realfast} 
and V-FASTR \citep{vfastr_wayth2011}, and are being used to analyze the efficiency of the respective detection pipelines.


\cite{Farah2019MNRAS} injected 2,000 mock FRBs via the real-time injection system at UTMOST (see Sec 3.4 in \citealp{Farah2019MNRAS}) and found that their pipelines successfully recover around 90\% of the injected mock FRBs.

In this paper, we provide full details of the mock FRB injection system implemented at UTMOST, and analyse the results of injecting 20,000 mock FRBs into live data streams, performed during standard pulsar timing and FRB search operations at the facility. These mock FRB injection tests have allowed us to investigate the performance of our pipeline in depth, by breaking down the recovery fraction at every step of the search and identifying the reasons for the missing 10\% of the injected mock FRBs reported in \cite{Farah2019MNRAS}. Additionally, we probe the pipeline's efficiency over a much broader parameter space in pulse DM, SNR and width. 

The paper is organised as follows: we briefly describe the observing and FRB detection system at UTMOST in Section~\ref{sec:UTMOST}. The full details of the live mock FRB injection system are presented in Section~\ref{sec:Mock FRBs} and the methods employed for the current study in Section \ref{sec:Methods}. The results and analyses of the the recovered mock FRBs are in Section~\ref{sec:Results}. We draw particular conclusions about FRB injection at UTMOST, and advocate for injection to be implemented at all FRB search facilities in Section~\ref{sec:Conclusions}.

\section{UTMOST}
\label{sec:UTMOST}

The Molonglo radio telescope is a 1.6 km long radio interferometer located 40 km east of Canberra in Australia. Composed of 352 individual elements (called modules) lined up in the East-West direction, it has a total collecting area of 18,000 m$^2$ which is the largest in southern hemisphere.
Built in the 1960s \citep{MillsCross1963}, the telescope was given a new life with the start of the UTMOST project in 2013 \citep{utmost} which installed a new GPU based software correlator as a part of the digital backend, transforming the telescope into an FRB-finding instrument (see \cite{utmost} for details). UTMOST has a bandwidth of 31.25~\rm{MHz} (820$-$851.2 \rm{MHz}) and an elliptical Field of View (FOV) of approximately $4^{\circ} \times 2.8^{\circ}$ which is uniformly tiled using 351 narrow fan-beams created by phasing the array to different sky positions within the primary beam.
We briefly summarise UTMOST's FRB search pipeline below, but refer the readers to \cite{Farah2019MNRAS} and \cite{utmost} for more detailed descriptions.

\subsection{The FRB detection pipeline}
\label{subsec:FRB detection pipeline}

Digitised, complex sampled, and channelised data from individual modules are searched for bright impulsive RFI in real-time and replaced with pseudo-random noise before beamforming. Individual fan-beams are then square-law detected, averaged to 327.68~$\rm{\upmu s}$ and 97.65~$\rm{kHz}$ resolution, and normalised to a mean of 0 and variance of unity (measured for the first 30 seconds of data on a per channel basis), before being searched for FRBs.
UTMOST's FRB detection pipeline uses a multi-beam version of \textsc{HEIMDALL}\footnote{\href{multi-beam_HEIMDALL}{https://github.com/ajameson/heimdall\_multibeam}}, a GPU-based dedispersion and accelerated search algorithm developed by \cite{Barsdell_PhD}. Besides UTMOST, \textsc{HEIMDALL}\footnote{\href{HEIMDALL}{https://sourceforge.net/p/heimdall-astro/wiki/Home/}} has also been used at Parkes \citep{SUPERB, Champion2015, Pravir_repeater, FRB180301_D_Price}, DSA-10 \citep{DSA-10}, GBT \citep{Aggarwal2020_GREENBURST}, SRT \citep{Sardinia_Pilia}, FAST \citep{FAST_Drift_scan} and STARE2 \citep{STARE2}, accounting for a significant fraction of all the published FRBs to date.

\textsc{HEIMDALL} performs a boxcar search for FRBs in the specified DM and width range. The spacing of trials in DM are set by a configurable tolerance for maximum loss in SNR (i.e. DM-tolerance) between DM trials and the boxcar widths are logarithmically spaced, spanning the range from 0 to  $2^{n}$ samples, where $n$ is a user-specified parameter. At UTMOST, our search range in DM is 0-5,000~\dm, and the DM-tolerance has been set at 1.20 with $n$ being set at 8, i.e. the widest boxcar we search is 256 samples or 83.8 \rm{ms} wide.

Due to the presence of 4 telecommunication bands within the observing band, impulsive RFI events, usually generated by mobile handsets, are detected by \textsc{HEIMDALL} as potential FRBs. 
To deal with the large number of false candidates, we have implemented a machine learning classifier based on the Random Forest algorithm \citep{Farah2019MNRAS} which extracts features from the raw filterbank data for each candidate in the list produced by \textsc{HEIMDALL} and assigns a classification probability of the signal being astrophysical. 
In order to keep the total number of candidates manageable, the pipeline filters the candidates before handing them to the Random Forest classifier. 
All candidates with SNR $<9$, boxcar width $>128$ samples (41.9 ms) or DM $< 50$~\dm are rejected. Candidates with SNR $<9$ or width $>128$ are unlikely to be of astrophysical origin, while those with DM $<50$ are expected to be galactic sources. The remaining candidates are processed by the classifier and if the evaluated probability is above 0.6, capture of raw-voltage data to disk at the native resolution of the instrument is triggered and a notification e-mail is sent out to the observers.

\subsection{The UTMOST scheduler}
\label{subsec:observing}

Observations at UTMOST are carried out almost completely autonomously using a bespoke scheduler as part of the Survey for Magnetars, Intermittent pulsars, RRATs and FRBs (SMIRF). 
The mock FRB injections at UTMOST have been designed to be scheduled by the SMIRF observation scheduler and therefore we briefly introduce its functionality in this section. A complete description of the SMIRF scheduler and its features can be found in \cite{SMIRF}. 
SMIRF typically uses a standard list of sources (including pulsars, FRBs, magnetars, RRATs) with specified target cadences to schedule observations automatically. Since UTMOST can only observe in transit mode, the time to slew to the target coordinates and the duration for which the source stays within the primary beam are also factored in while choosing the next source to observe and the length of the observation.
At the start of the observation, SMIRF sends all chosen parameters to the observing backend in an XML packet which is parsed and then implemented by the telescope control software to carry out the observation.
The ability to schedule observations dynamically enables UTMOST to time $\sim$400 pulsars every month while searching for FRBs commensally, with minimal human supervision.

\section{The mock FRB injection system}
\label{sec:Mock FRBs}
Mock FRB signals could be injected at various stages, such as in the digital backend onto the beamformed intensity data, into the individual antenna voltage streams prior to beamforming, or even electromagnetically into the antenna feeds prior to digitization. 
In general, injection at early stages of data capture and processing provides a more comprehensive test of the instrument's FRB detection efficiency. 
However, for a telescope with large number of individual elements and with a large median Fresnel length ($>$ 10,000 km) as UTMOST does, electromagnetic injection through artificial signals in the field or voltage-stage injection in data stream from every element becomes impractical, leaving power injection into the filterbanks at the digital backend stage as the only realistic option.

The mock FRB injection system at UTMOST is implemented to inject mock FRBs into beamformed fan-beam data. The injection system has two main parts: a) mock FRB generation, and b) mock FRB live injection. 

\subsection{Mock FRB generation}
\label{subsec:Furby generation}

Mock FRBs are generated offline and stored in a database prior to any injections. Before the FRB signal is simulated, the maximum number of samples required to accurately represent the final dispersed FRB must be computed. For UTMOST's bandwidth, time resolution and maximum DM searched (5,000 \dm), this value is fixed at 9,000 samples, but is a configurable parameter which can be modified before generating a new mock FRB database.

Each mock FRB block is a 2-dimensional noise-free frequency-time profile. To generate a mock FRB with a desired SNR, DM and width, we adopt the following steps:
\begin{enumerate}
    \item We generate a time series with a simple Gaussian signal of unit height and width in each channel. The signal is generated at 32 times the desired time resolution (and 10 times the frequency resolution during the dispersion step in \ref{Dispersion}) and is averaged down after applying the effects of scattering and dispersion (as explained in \ref{scattering} and \ref{Dispersion}). This ensures that we get an accurate representation of even the narrowest of our mock FRBs, and allows for sub-sample shifts during dispersion. We do not add any noise to our template yet.
    
    \item To generate the required SNR for the event, we need prior knowledge of noise in the live data stream into which the mock FRB would be added. By design, the mock FRBs are injected into beam-formed intensity data of individual fan-beams (also see section \ref{subsec:Furby injection}) where noise has been normalised to 0 mean and unit variance in each frequency channel. We assume ideal thermal noise statistics and scale our profile to the desired area (fluence) and width, such that when added onto the live data, our mock FRB would expectantly have the desired SNR. However, RFI in the live data that leads to a non-Gaussian distribution of noise samples may mean that the noise is higher, and hence mock FRB SNR lower.

    \item We then modulate the amplitude of our mock FRB to simulate a spectrum with different spectral indices of a power law. For the small fractional bandwidth of UTMOST, a power law can be approximated with a linear relation. We generate a list of scale factors for each frequency channel based on a randomly chosen slope of the straight line. The value of the slope is sampled from a uniform random distribution between $-0.5$ and 0.5, approximating a spectral power law index ranging between $-18$ and 18. The scale factors are chosen appropriately to conserve the total signal across all channels in the mock FRB, and then applied to the FRB block.
    
    \item
    \label{scattering}
    Next, we convolve our mock FRB profile in each channel with an exponential scattering kernel. The scattering time scale $\tau$(ms) at 843 MHz is chosen based on the relation:
    \begin{equation}
        \tau = \lvert~ \textrm{DM} / 1000 + R ~\rvert
    \end{equation}
    where $R$ represents a random number drawn from a normal distribution with a mean of 0 and standard deviation of 2. 
    This definition allows for a spread in the values of $\tau$ while keeping it proportional to the requested DM of the FRB, i.e., high-DM mock FRBs are likely to have longer scattering times and vice versa. Following a thin screen model of turbulence, the scattering time in each channel is made frequency ($\nu$) dependent and scales as $\nu^{-4}$.
    
    \item The frequency channels of the FRB are then scaled by the UTMOST bandpass. This bandpass has been  measured on bright, broadband non-scintillating sources (e.g. the pulsar PSR J1644$-$4559) and verified with continuum calibration sources. This step is necessary because the detection pipeline assumes that every signal would be affected by our bandpass and optimises for it before searching for signals. It is ensured that the total signal of the mock FRB stays conserved after each channel is multiplied with its corresponding bandpass weight.
    
    \item 
    \label{Dispersion}
    In order to simulate the effect of DM smearing across finite channels, we further channelise our mock FRB by a factor of 10, by creating 9 additional copies of each channel. Each subchannel is then shifted (rolled), with respect to the centre channel, by a delay equal to the dispersive delay at that frequency for the requested DM of the mock FRB. Finally, every 10 adjacent subchannels are averaged back into a single channel. This disperses our mock FRB profile while including the effect of intra-channel dispersion smearing.
    
    \item Finally, the mock FRB block is associated with a unique ID and saved on disk in a database. The values in the mock FRB block are floating point numbers and no loss due to quantisation takes place. A catalogue of the generated IDs is simultaneously maintained inside the database, which is used in the injection process, as explained in section \ref{subsec:Furby injection}. 
    
\end{enumerate}

The code for generating a mock FRB database is publicly available on GitHub\footnote{\href{Furby}{https://github.com/vg2691994/Furby}}.
\subsection{Estimating injected SNR}

\label{Injected SNR}

Because we simulate both the DM dispersion in finite frequency channels, and scattering for each mock FRB, the final SNR and width, after injection into live data, can differ from the initial SNR and width given as input to the generator. 


To derive an accurate estimate of the injected SNR and width, we add Gaussian noise to the mock FRB block and generate a dedispersed time series (averaged along frequency axis). 
We run a box-car convolution kernel on the dedispersed time series and find the maxima of the convolution, which is divided by the standard deviation of the noise in the off-pulse region. This value is recorded as the \textbf{Injected SNR} and the box-car width at which the maxima occurs is recorded as the \textbf{Injected width}. 
The box-car convolution is not required to run in real-time allowing for a rigorous search. Our box-car width trials range from 0 to 100 ms wide, incrementing in steps of one sample (327.68 ${\upmu}$s). This makes our measured value of the injected SNR equivalent to the value of ``maximum recoverable SNR" by a box-car search algorithm, providing a useful metric to compare against the performance of practical implementations of the algorithm like in \textsc{HEIMDALL}.

The Injected SNR, Injected width, and Injected DM values are added to the catalogue of simulated FRBs and made available for post injection analysis.

We note that implicit to this method of evaluating SNR is a shape mismatch problem between the pulse profile and the box-car shaped template. A matched-filter convolution would be a more optimal way of estimating the true injected SNR. 
However, due to its low computational cost, the box-car convolution search has become a standard practice among the pulsar/FRB community and the box-car SNR is widely used as the detection significance of transient signals.
Additionally, for the purpose of analysing the performance of \textsc{HEIMDALL} we consider the maximum recoverable SNR in a box-car convolution to be a better metric for comparison than the SNR derived using the optimal-matched-filter approach.
\subsection{Mock FRB injection}
\label{subsec:Furby injection}

\begin{figure*}
    \centering
    \includegraphics[width = 0.99 \textwidth]{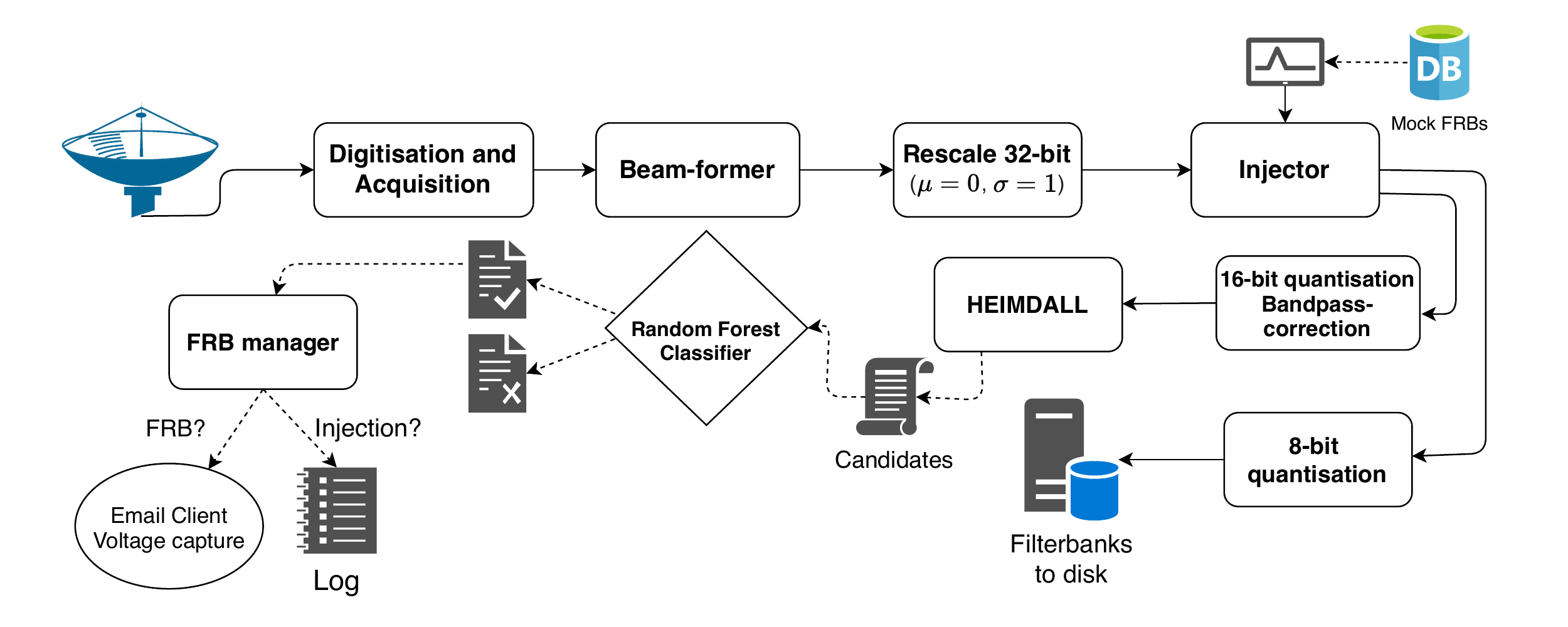}
    \caption{Signal processing and injection pipeline at UTMOST. Solid lines indicate flow of live data and dashed lines indicate flow of packets of information. Full details are given in Section \ref{sec:Mock FRBs}.}
    \label{fig:pipeline}
\end{figure*}

Once the mock FRBs have been generated, the injection process is simply the addition of the mock FRB block into the live data in the desired fan-beam at the desired time. 
To achieve this, we have introduced an additional process, called the ``injector'', in our signal processing pipeline. 
The injector operates on the beam-formed intensity data after it has been re-scaled and normalised to 0 mean and unit variance statistics. 
At this stage, the data have been integrated to 327.68~$\rm{\upmu}$s and 97.65 \rm{kHz} resolution.
Our pre-generated mock FRBs match this resolution such that adding them to the data-stream does not require re-scaling or interpolation.
Figure \ref{fig:pipeline} shows the flow of data through the signal processing pipeline at UTMOST highlighting the relevant parts for injection. A more detailed description of other parts of the pipeline can be found in \cite{utmost} and \cite{Farah2019MNRAS}.

To trigger injections in an observation, we have made modifications to the SMIRF's observation scheduler. While scheduling the next observation, SMIRF reads the catalogue of available mock FRBs in the database and selects $N$ IDs randomly. The value of $N$ is set in the configuration file for SMIRF.
Then, based on the number of beams and length of the observation, it generates an injection beam and an injection timestamp for each of the chosen IDs. Finally, it attaches all $N$ sets of ID, beam and time-stamp to the XML packet it sends to the observing backend to start the observation.

The observing backend passes the information about the scheduled injections to the injector process at the start of the observation. The injector reads the requested injection IDs, beams and timestamps and loads all the mock FRB blocks in memory. 
Once the live data starts flowing, it keeps counting samples until it sees the injection timestamp, and then adds the mock FRB block to the live data-stream in the requested beam. At the same time, it saves a snippet of live data (with the mock FRB added) to disk. This copy is saved only to aid in the analysis of the injected mock FRB later and is not an essential component of the pipeline.

\subsection{Mock FRB detection}
\label{subsec:Furby detection}

Apart from the mock FRB injector, the entire search pipeline remains blind to mock FRBs, treating them as real astrophysical signals.  The list of FRB candidates produced by \textsc{HEIMDALL} is ingested by the machine learning classifier and it assigns a probability to every candidate that passes through its SNR and width thresholds (SNR $\geq$ 9 and box-car width $\leq$ $2^{7}$ samples) exactly as would happen with no injections.
If a candidate is classified as a potential FRB (or single pulsar pulse), it is sent to the FRB manager for issuing e-mail alerts to observers and triggering a voltage capture. 
However, the FRB manager cross-matches these candidates with the list of mock injections in the current observation -- if the detected candidate has the same beam and timestamp as a requested injection, the candidate is classified as a mock FRB and silently logged to a file without issuing alerts or voltage capture triggers.

\section{Methods}
\label{sec:Methods}

\begin{figure*}
    \centering
    \includegraphics[width = 0.325\textwidth]{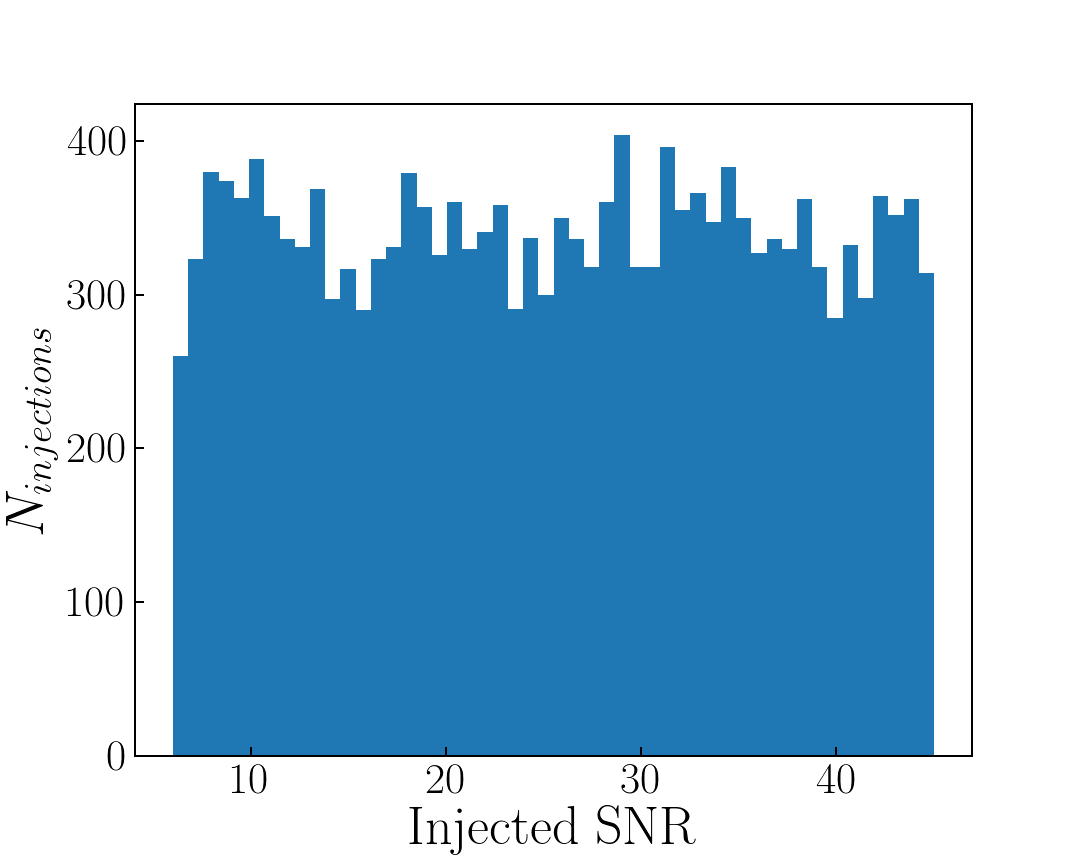}
    \vspace{0.0mm}
    \includegraphics[width = 0.325\textwidth]{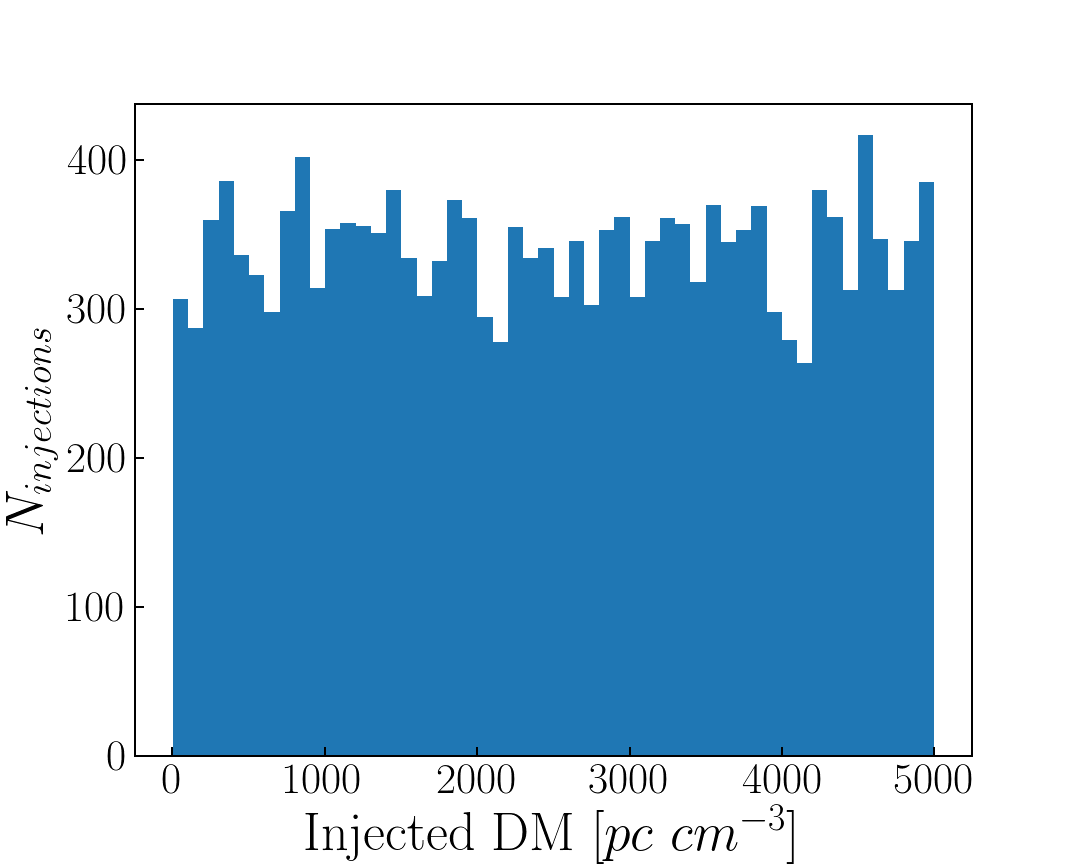}
    \vspace{0.0mm}
    \includegraphics[width = 0.325\textwidth]{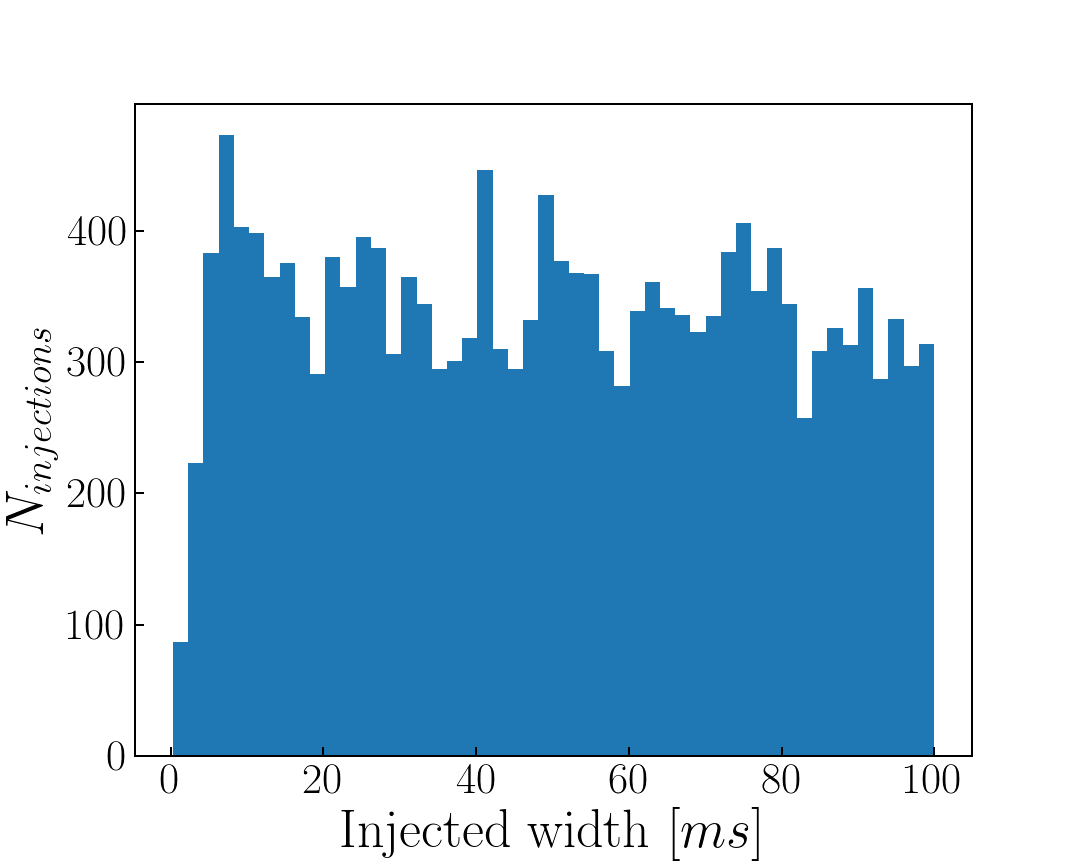}
    
    \caption{Histograms of the injected mock FRB parameters showing uniform sampling in SNR range (6, 45), DM range (5, 5000)$~$\dm, width range (0.01, 100) $\rm{ms}$.}
    \label{fig:Injection statistics}
\end{figure*}

\begin{figure*}
    \centering
    \includegraphics[width = 0.325\textwidth]{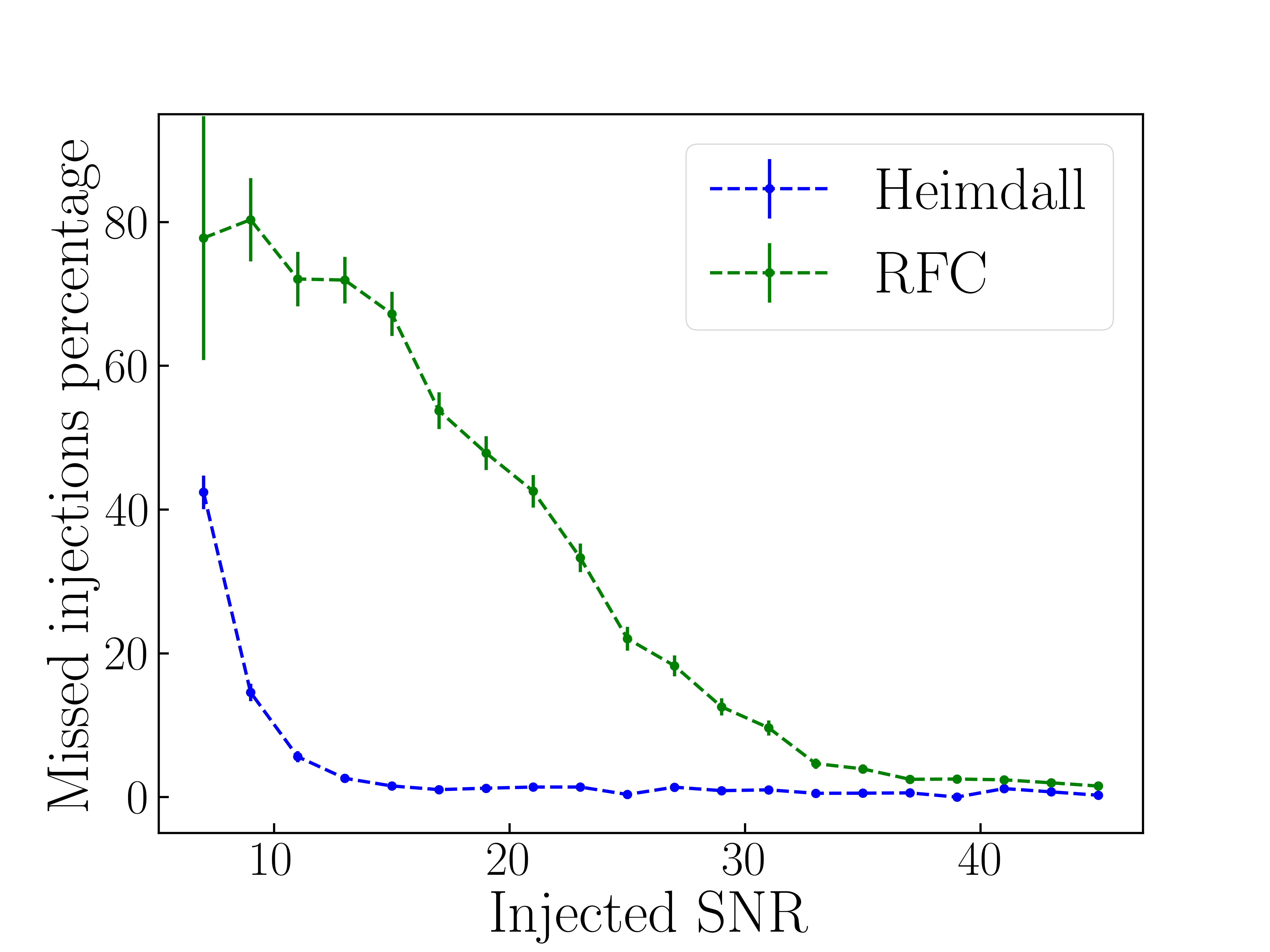}
    \vspace{0.0mm}
    \includegraphics[width = 0.325\textwidth]{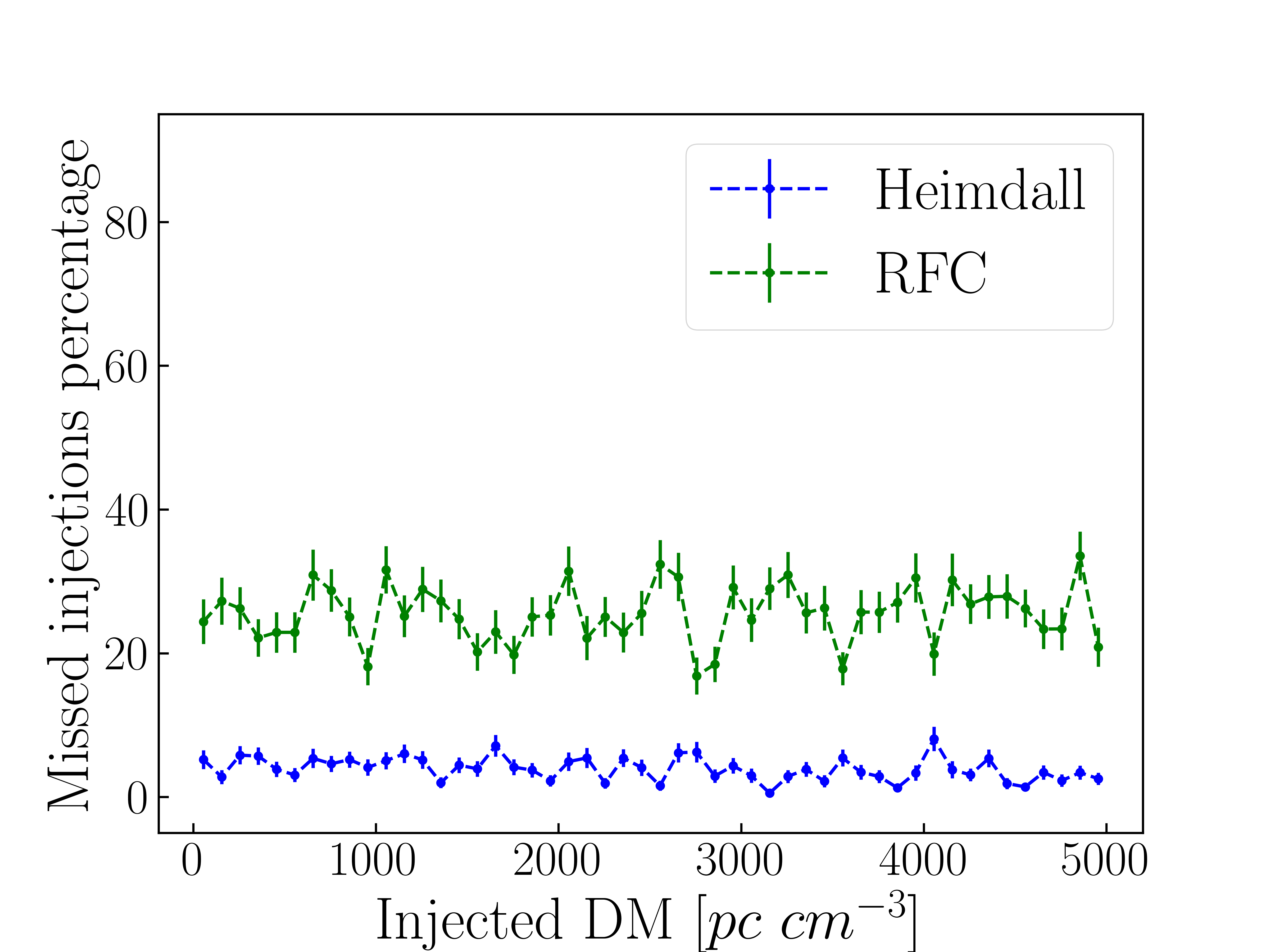}
    \vspace{0.0mm}
    \includegraphics[width = 0.325\textwidth]{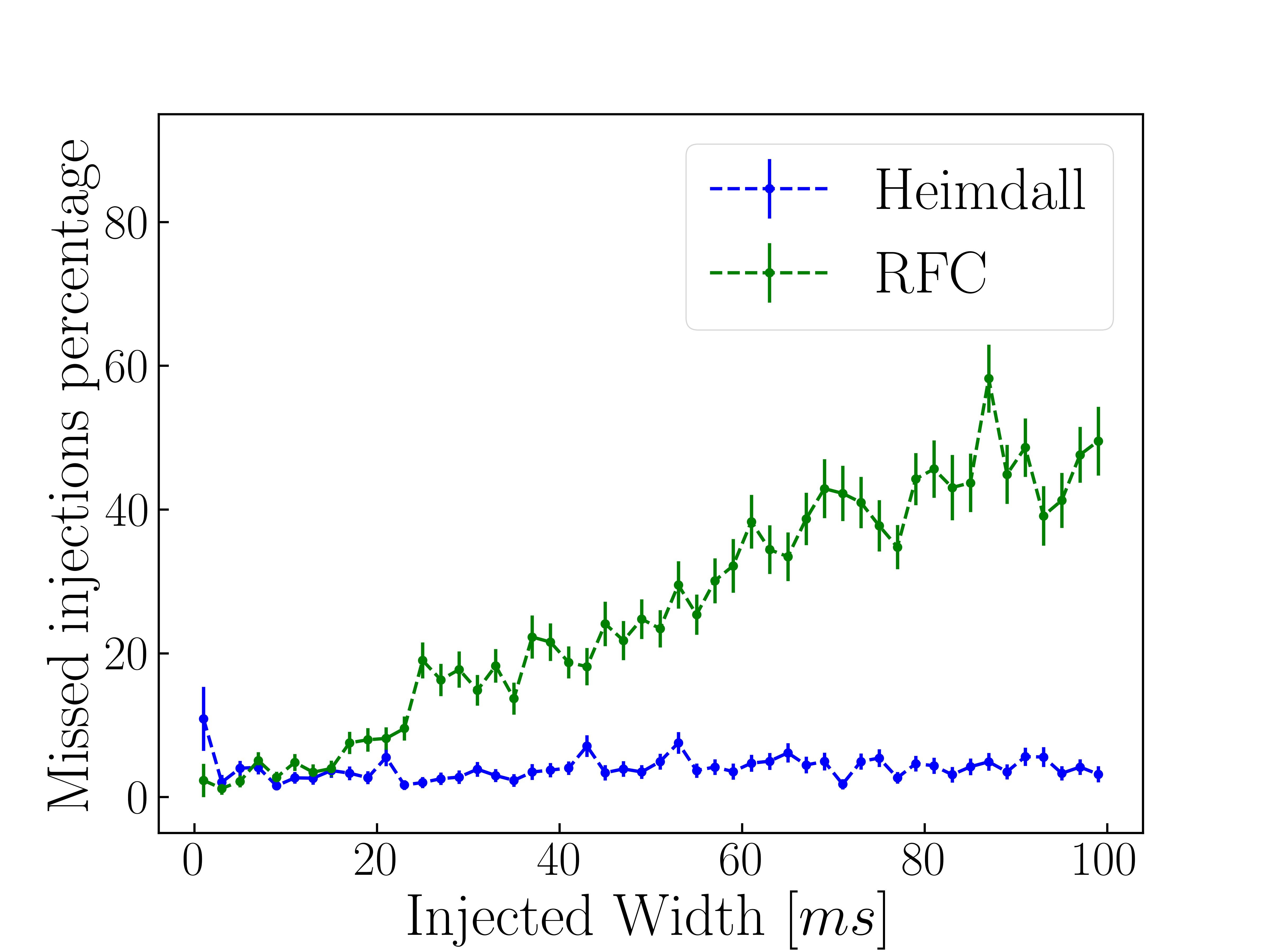}
    
    \caption{Trends in percentage of missed injections by \textsc{HEIMDALL} and the Random Forest Classifier (RFC) as a function of the injected SNR, DM and width. While the high fraction of missed injections by the RFC at low SNRs is concerning at first sight, it is a consequence of marginalising over the higher missed fraction of large number of mock FRBs with wider widths (see Section \ref{sec: Recovery stats}).}
    \label{fig:Missed_fractions}
\end{figure*}

\begin{figure*}

 \begin{subfigure}[b]{.304\textwidth}
    \centering
    \includegraphics[width = \textwidth]{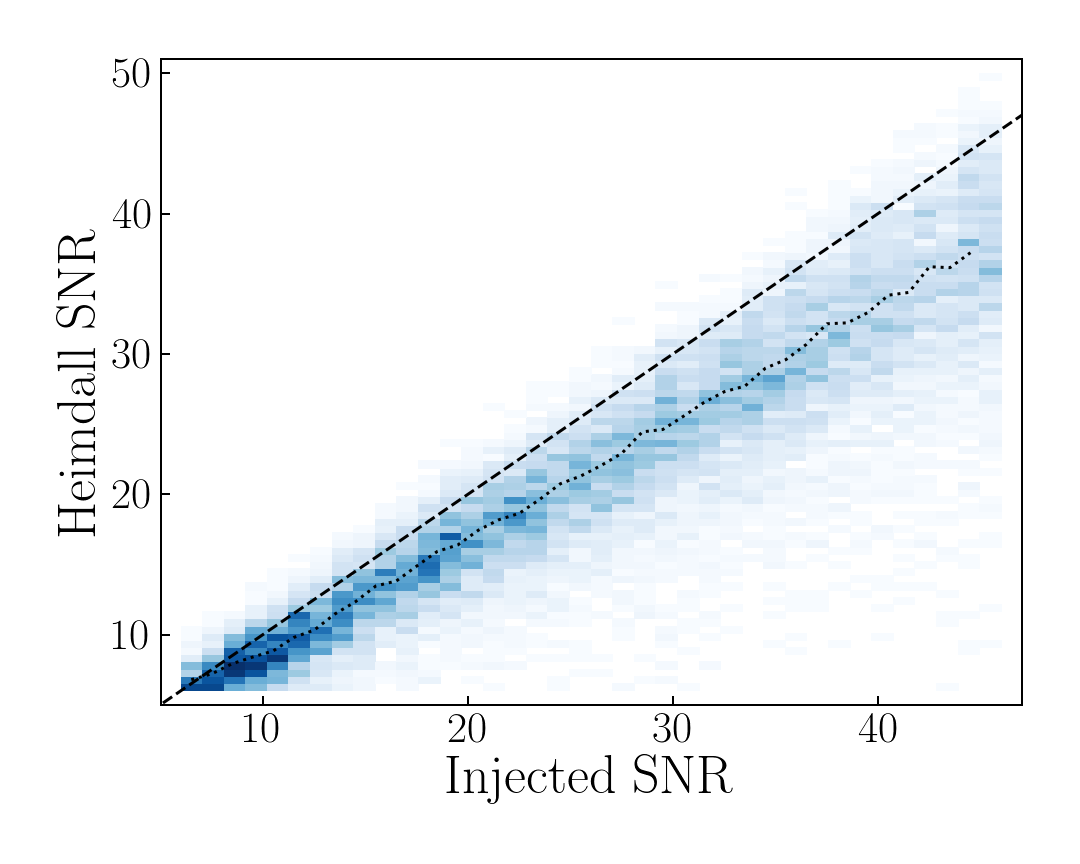}
    \caption{SNR recovery}
    \label{fig:HEIMDALL SNR recovery}
 
 \end{subfigure}
 \hspace{0.1em}
 \begin{subfigure}[b]{.305\textwidth}
    \centering
    \includegraphics[width=
    \textwidth]{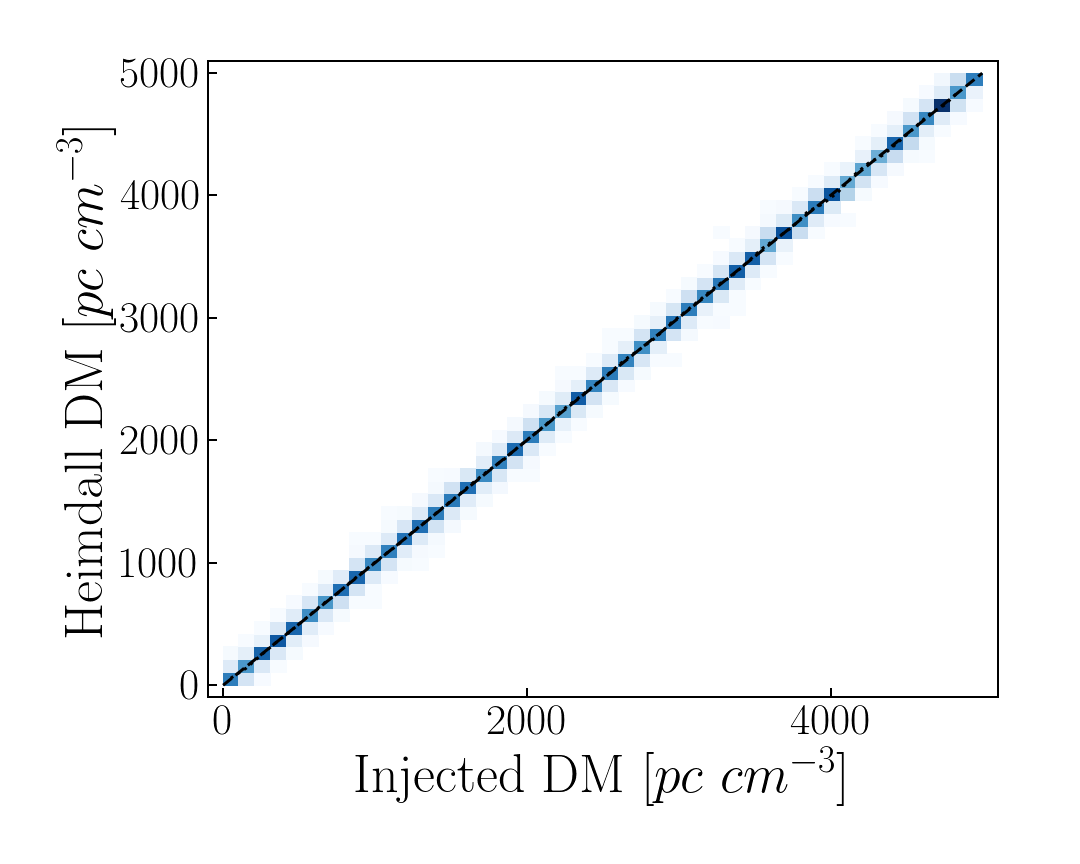}
    \caption{DM recovery}
    \label{fig:HEIMDALL DM recovery}
 \end{subfigure}
 \hspace{0.1em}
 \begin{subfigure}[b]{.37\textwidth}
    \centering
    \includegraphics[width=\textwidth]{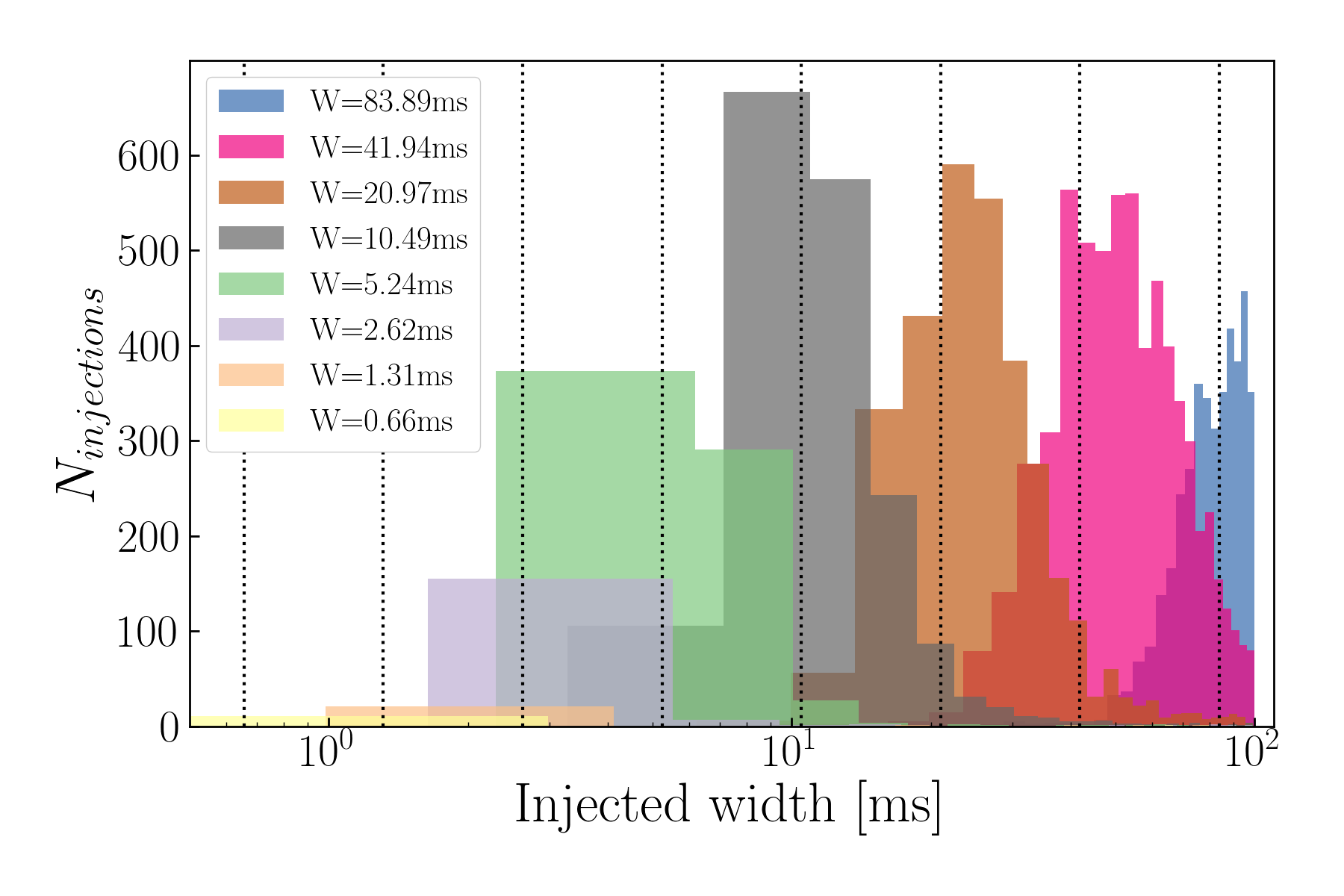}
    \caption{Width recovery}
    \label{fig:HEIMDALL width recovery}
 \end{subfigure}
    
    \caption{Comparison of SNR, DM and width values recovered by \textsc{HEIMDALL} versus the corresponding injected values. Panels (a) and (b) show the density distribution of recovered values of SNR and DM against the injected values. The black dashed lines represent the lines of 1:1 ratio. The black dotted line in panel (a) shows the running median of the distribution with bin size of 1 SNR unit. In panel (c) we show the histograms of injected widths of mock FRBs detected in a given boxcar trial. The vertical dashed lines indicate the boxcar trial widths searched by \textsc{HEIMDALL}. The discrepancy of approximately 20\% between the injected and recovered SNR is clearly seen in (a), and discussed in depth in Section \ref{sec: Analysis.HEIMDALL}. Panel (b) highlights that the DM is recovered with high fidelity. Panel (c) shows that for the most part, FRBs are recovered with the nearest matching search width, although with considerable scatter (see Section \ref{sec: Analysis.HEIMDALL}).}
    \label{fig:Heimdall recovery}
\end{figure*}

\begin{figure*}
    \centering
    \includegraphics[width = 0.45\textwidth]{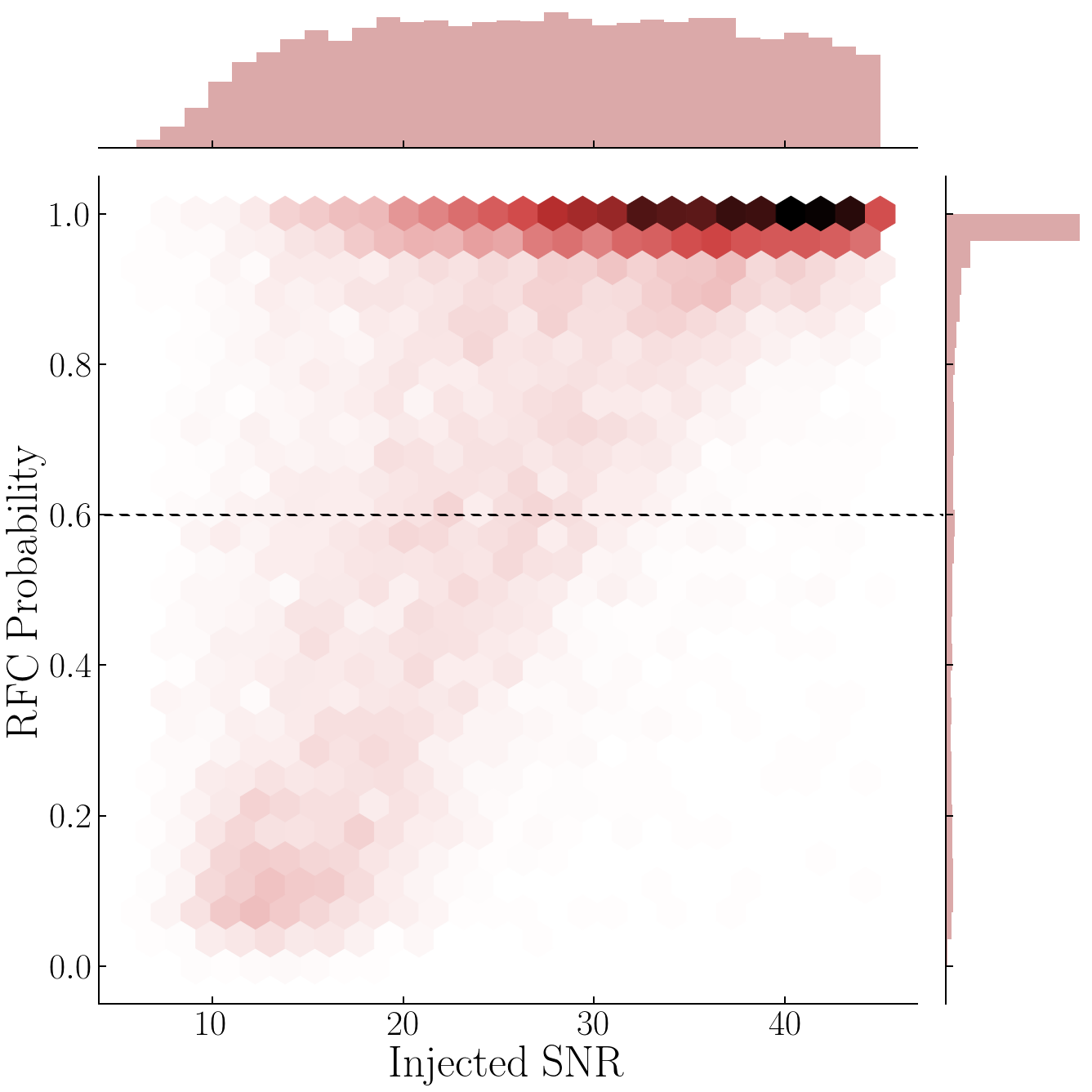}
    \vspace{0.0mm}
    \includegraphics[width = 0.45\textwidth]{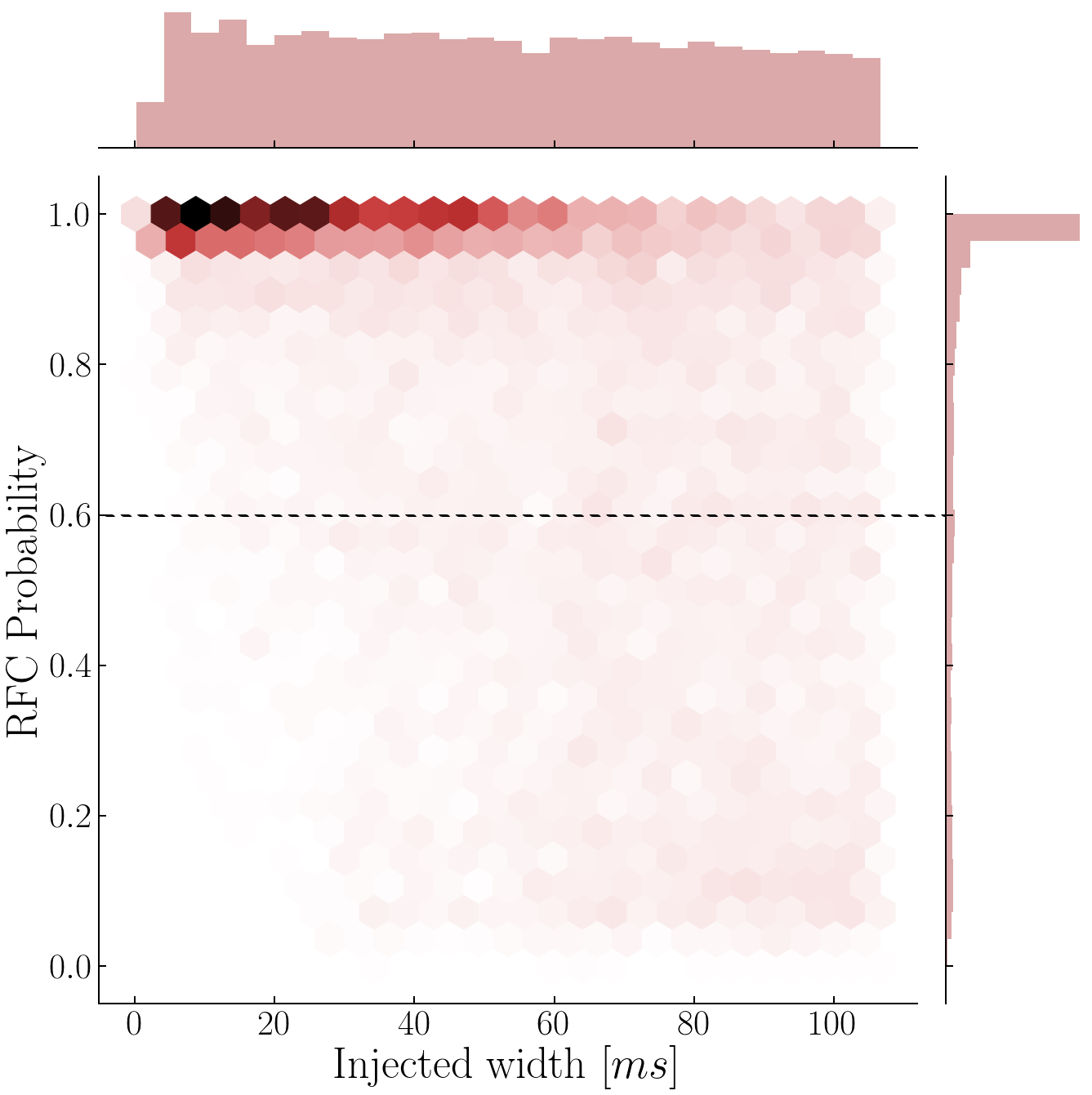}
    \caption{Density distribution of the classification probability evaluated by the Random Forest classifier (RFC) for the injected mock FRBs as a function of the Injected SNR and Injected Width. Horizontal dashed line at RFC probability $= 0.6$ indicates the current threshold for classification of a signal as astrophysical. The colours scale linearly with probability density in the plot.}
    \label{fig:RCF prob}
\end{figure*}

We injected 20,000 mock FRBs in our live data-stream over 6 days of continuous observations in mid-September 2019. We expected the recovery rate of our pipelines to be dependent upon the injection parameters like width and SNR, and therefore that the overall recovery efficiency would also depend upon the distribution of these parameters. 
Keeping in mind that the goal of this study was to characterise the performance of the pipelines and not computing the FRB detection efficiency of our survey, we chose to keep a uniform distribution in SNR, width and DM parameter space.
We uniformly sampled the SNR range from 6 to 45, a DM range from $5~$\dm$~$to $5,000~$\dm$~$and a width range from $0.01~$\rm{ms} to $100~$\rm{ms}. However, the broadening effect of intra-channel dispersion smearing disproportionately affects the narrowest of our mock FRBs, resulting in a lack of injections at narrow end of our width distribution. The resulting distributions of SNR, DM and width of the injected mock FRBs are shown in Figure \ref{fig:Injection statistics}.

For every injection for which \textsc{HEIMDALL} reports a candidate in the same beam, within 100 \rm{ms} of the injection time-stamp, and at a DM within $\pm100~$\dm of the injected DM, the mock FRB is counted as detected by \textsc{HEIMDALL}. If any of these conditions is not met we count it as missed.

The list of candidates produced by \textsc{HEIMDALL} is filtered for SNR ($\geq9$) and box-car width ($\leq2^{7}$samples $\approx41.9~\rm{ms}$) and then passed to the Random Forest Classifier (RFC). For every candidate the RFC evaluates the probability of the candidate being an FRB. If an injected mock FRB is assigned a probability of $\geq0.6$, it is counted as detected, and as missed otherwise. 


During real-time processing, the RFC excludes candidates above a box-car width of 41.9 ms because the number of candidates becomes prohibitive in our RFI environment. Our injected mock FRBs have widths up to $100~\rm{ms}$, and unfortunately this box-car threshold width meant that roughly half of our injected mock FRBs were not evaluated by the RFC running on live data. Excluding candidates above this threshold is necessary to achieve real-time processing, but somewhat hinders the goals of this study. 

Consequently, during post processing, we use the snippets of live data $+$ mock FRB, saved by the injector, to evaluate the RFC probability of \textsc{HEIMDALL} candidates with box-car width $> 41.9$ ms and SNR $< 9$ down to the detection threshold of \textsc{HEIMDALL} at 6-sigma. We still exclude those injections which are missed by \textsc{HEIMDALL}, despite having their injected data snippets available on disk.

\section{Results and Analysis}
\label{sec:Results}

\begin{figure}
    \includegraphics[width = 0.42\textwidth]{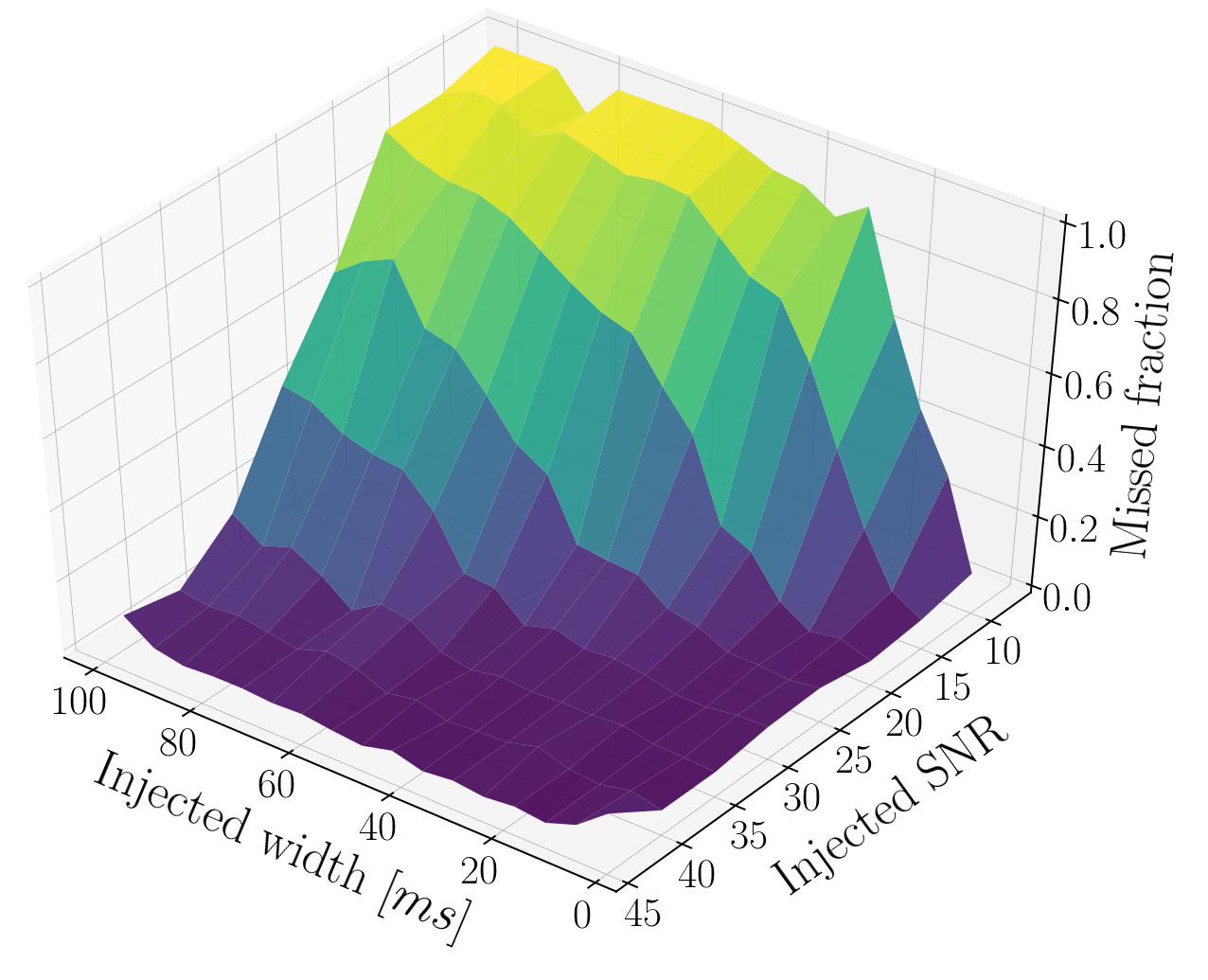}
    \caption{The fraction of injected FRBs missed by the Random Forest Classifier shown as a function of SNR and width. The majority of the missed injections are those with large widths and low SNR. The narrower width FRBs are recovered effectively at all SNRs.}
    \label{fig:SNR-w-3d}
\end{figure}
\subsection{FRB recovery fractions}
\label{sec: Recovery stats}

Marginalising over all parameters \textsc{HEIMDALL} detects 96\% of the injected mock FRBs and the RFC correctly classifies 75\% of those detected by \textsc{HEIMDALL}.
The number of injections missed by \textsc{HEIMDALL} are plotted as a percentage of total injected, against the Injected SNR, DM and width in Figure \ref{fig:Missed_fractions}. 

\textsc{HEIMDALL} preferentially misses low SNR injections but shows little dependence on the injected DM and width. The recovery fraction remains greater than 90\% for injections above the detection threshold of our surveys (SNR $\geq$ 9), and is a reassuring indicator of the performance of \textsc{HEIMDALL} over a broad range of testing, even in an RFI-affected environment.

The number of injections missed by the Random Forest Classifier as a percentage of total ingested (detected by \textsc{HEIMDALL}) is plotted against the injected SNR, DM and width in Figure \ref{fig:Missed_fractions}. 
The RFC shows strong dependence on the injected SNR and the injected width, while its performance is uniform across the tested DM range.
The RFC extracts the features from filterbanks after the data have been dedispersed at the candidate's DM, hence the performance independence from DM is along expected lines.

However, the percentage of missed injections at low injected SNR seems very large ($\approx$ 80$\%$) and potentially alarming. This can be understood as a consequence of marginalising over the large range of injected widths sampled uniformly up to 100 \textrm{ms}.
The bulk of our mock FRBs have injected widths larger than 20 \textrm{ms}, and the RFC has a high missed fraction for these high widths. 
This is driving the overall fraction of missed FRBs by the RFC and resulting in seemingly alarming levels of missed injections at lower SNRs.
We discuss this in more detail in section \ref{sec: Analysis.RFC}.

\subsection{Analysis of mock FRB recovery fractions}
\label{sec:Analysis}

\subsubsection{\textsc{HEIMDALL}}
\label{sec: Analysis.HEIMDALL}

To better understand the performance of \textsc{HEIMDALL}, we compare the values of SNR, DM and width recovered by \textsc{HEIMDALL} against the injected values, as shown in Figure \ref{fig:Heimdall recovery}. 

For any given injected FRB, the recovered SNR will be affected by the mismatch between the actual FRB width and DM and the closest trial width and DM searched by \textsc{HEIMDALL}. However, we find that the detected SNR values by \textsc{HEIMDALL} are systematically lower than injected SNR values by $\sim$20\%, which is a larger drop that can be explained by the finite number of width and DM trials. 
The systematic decrease in the detected SNR by \textsc{HEIMDALL} implies that mock FRBs with low injected SNR preferentially drop below \textsc{HEIMDALL}'s detection threshold (SNR $=$ 6), resulting in the sharp increase in missed FRBs below an injected SNR of $\sim$9. We find that $\sim$0.1\% of injections showed a large deficit in their recovered SNR ($>75$\%). Analysis of the live data snippets saved for these injections show that they are typically heavily contaminated by RFI, accounting for the large losses in SNR.

The DM values are recovered accurately throughout the tested range. 
We investigate the width recovery accuracy of \textsc{HEIMDALL} by looking at the distribution of the injected widths of the mock FRBs which are detected in a given box-car width trial of \textsc{HEIMDALL}. 
\textsc{HEIMDALL} has width trials spaced by powers of 2 (see section \ref{subsec:FRB detection pipeline}), while our injected widths are distributed in a continuous range (see section \ref{sec:Methods}). 
As shown in Figure \ref{fig:Heimdall recovery}, the peak of the distributions of the injected widths match the widths at which they are detected.
However, we observe significant scatter and long tails in each distribution. $\sim$28\% of the injections are detected in the incorrect width trial, i.e. in a width trial which is off by greater than a factor of 1.5 from the box-car width which is closest to the injected width.
To explain this more clearly, let $W_{I}$ be the injected width of a mock FRB, and $W_{H}$ the box-car width at which it is detected by \textsc{HEIMDALL}. If $n$ is the number of injections where $W_{I} / W_{H} < 0.75$ or $W_{I} / W_{H} >1.5$, and $N$ is the total number of injections detected by \textsc{HEIMDALL}, then, we find that the fraction $n / N$ is equal to $\sim$ 0.28.

To understand the reason behind the loss in SNR and reduction in the recovered width values, we investigate \textsc{HEIMDALL}'s internal functions used to determine the SNR and width of a candidate.
We find that a sub-optimal implementation of the baseline normalisation method inside \textsc{HEIMDALL} is causing an overestimation of the noise rms in the data in the presence of RFI. 
This overestimation gets larger at higher width trials, biasing the recovered SNR and width to lower values than injected.
Specific details of the problem, and a possible solution are provided in Appendix \ref{appndix:A}.


\subsubsection{Random Forest Classifier}
\label{sec: Analysis.RFC}

We can understand the variation in performance of the RFC by looking at its evaluated probabilities for each injection, as a function of the injected SNR and width. Shown in Figure \ref{fig:RCF prob}, we find that the classification probability seems to decrease smoothly with decreasing SNR and as more and more injections cross the detection threshold of probability 0.6, we get a smooth increase in missed percentage of injections. 
This is partly justified by the fact that as the SNR decreases, the strength of the features describing an FRB also decrease, and hence the probability of a candidate being real starts to decline.
However, the fact that this decrease starts to occur even for candidates with SNR above 20 suggests that our RFC is too reliant upon features extracted from the intensity in the on-pulse region and lacks strong discriminatory power from features which are relatively independent of the total intensity of the burst, for example, the frequency modulation index, averaged over the candidate width. \cite{Farah2019MNRAS} give a full description of the features currently used by the RFC.

Similarly, Figure \ref{fig:RCF prob} also shows us the gradual decline in the RFC probability with increasing width of the injected FRB. This is also expected as wider pulses will have their power spread across multiple time bins weakening some of the features used by the classifier to make a decision.
However, another reason for the decline in probability with increasing width could simply be the lack of good training data set for the model. The RFC was trained using single pulses from bright pulsars \citep{Farah2019MNRAS} and the widest single pulses in the training data-set were $\approx$ 30 \rm{ms} wide from J1644$-$4559.

Finally, the high percentage of missed injections at low SNR is a result of our choice of uniform sampling up to large widths of 100$\rm{ms}$ and marginalising over the width range when computing the missed fraction in a given SNR bin. 
Figure \ref{fig:SNR-w-3d} shows a combined distribution of the missed injections as a function of the injected SNR and width. 
At narrow widths even low SNR injections are reliably recovered by the RFC pipeline. It is the majority of the high width low SNR injections which are missed, driving up the overall missed fraction by the RFC. 
This does not indicate that UTMOST's FRB survey would be missing $\sim$80\% of FRBs at 10-sigma level. Quantifying the missed fraction of real FRBs for \textsc{HEIMDALL} and the RFC would require knowledge of the intrinsic width and fluence distribution of the FRB population, which is still not well understood. Assuming most FRBs have widths $<10$ ms, as is the case with 14 of 16 FRBs detected with UTMOST to date, then our test show that UTMOST recovers these FRBs with $>$98\% efficiency.

\subsection{Survey completeness}

\begin{figure}
    \centering
    \includegraphics[width = 0.48\textwidth]{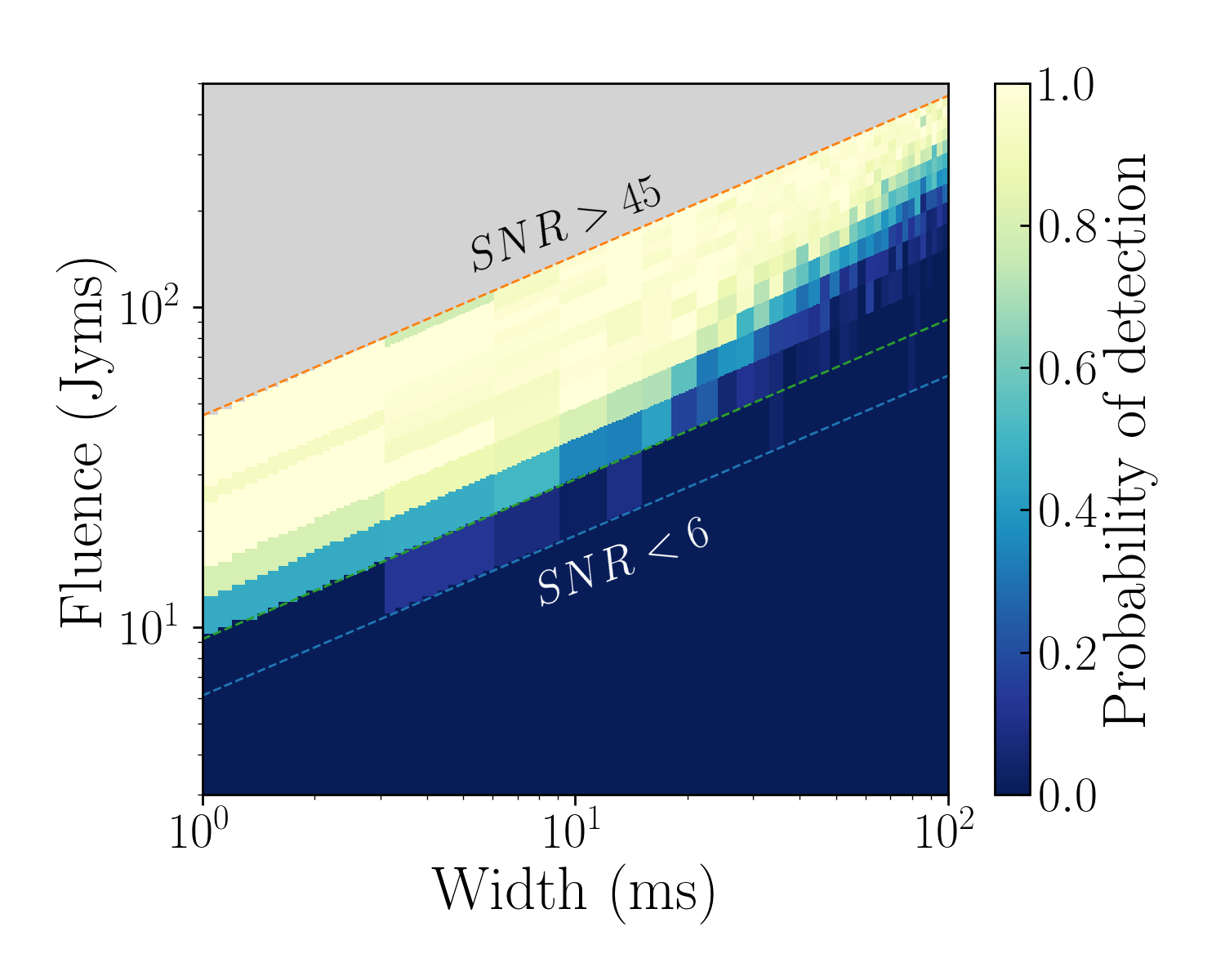}
    \caption{The completeness fraction of our survey to FRBs as a function of their observed fluence and detected width. Orange, green and blue dashed lines indicate lines of constant SNR of 45, 9 and 6 respectively. All FRBs lying in the SNR $<$6 region are below HEIMDALL's search threshold. The grey region above SNR $>$45 is not probed by the injection sample used in this paper, however, we expect our detection pipeline to recover those FRBs with high efficiency. The blue triangular region above the SNR of 9 shows the incompleteness of our survey to broad FRBs, as revealed by the injections.}
    \label{fig:Flux_width_completeness}
\end{figure}

The injections have shown that our detection pipeline is not 100\% complete in the parameter space probed by our survey. The correction factor for the completeness of the FRB survey described in \cite{Farah2019MNRAS} is presented as a global correction based on a set of 2,000 injections made prior to that study.
This study has shown that the correction factors develop quite strong dependence on pulse width in the much broader sample of FRBs analysed. We use the measured fraction of FRBs missed by our pipeline (shown in Figure \ref{fig:SNR-w-3d}) to estimate the completeness of our survey as a function of the observed fluences and detected widths of FRBs. This survey completeness fraction is plotted in Figure \ref{fig:Flux_width_completeness}. The recovery of mock FRBs has been found to be uniform across the tested range of DM, suggesting that our survey's completeness fraction is independent of the DM up to our search limit of 5,000 \dm.

We note that due to the finite time resolution (327.68 $\upmu$s) in our search data, FRBs with intrinsic widths less than the sampling time are also recovered at reduced SNR.  
The SNR drops by a factor of $\sqrt{w_{i}/\sqrt{(t_{s}^{2} + w_{i}^{2})}}$, where $w_{i}$ is the intrinsic width of the FRB (including the width broadening due to scattering) and $t_{s}$ is the time resolution of the data. Furthermore, for narrow FRBs with DM $\ga 250$ \dm, the finite frequency resolution (97.65 kHz) in our search data will cause the pulse to be smeared across multiple time samples, resulting in a decrease in the recovered SNR by the factor $\sqrt{w_{i}/\sqrt{(t_{DM}^{2} + t_{s}^{2} + w_{i}^{2})}}$, where $t_{DM}$ represents the smearing width due to the intra-channel dispersion. Consequently, our FRB detection pipeline misses a fraction of FRBs with narrow widths with a DM dependency. We need to take this effect into account while comprehensively calculating the completeness of our FRB survey.
However, this insensitivity to narrow FRBs is not due to deficiencies in \textrm{HEIMDALL} or the Random Forest Classifier, but is a function of the time and frequency resolution of the instrument, which are limited by the computational resources available for the survey. The resulting incompleteness in the survey can be accurately modelled analytically and does not require rigorous testing using the injections. Hence, we do not address the completeness fraction of our pipeline to narrow FRBs in this paper, but it will be included in a future paper describing the sample of FRBs detected in our current survey.

We note that we have not investigated the performance of the detection pipeline for spectral FRB properties such as the dispersion law index (currently fixed at $-2$), scattering spectral law index (currently fixed at $-4$), or patchy spectral features. The capacity to include these spectral properties already exists in the mock FRB generator code, and could be easily utilised for future studies.

\subsection{Additional Insights}
\label{sec: Additional insights}
\begin{figure}
    \includegraphics[width = 0.48\textwidth]{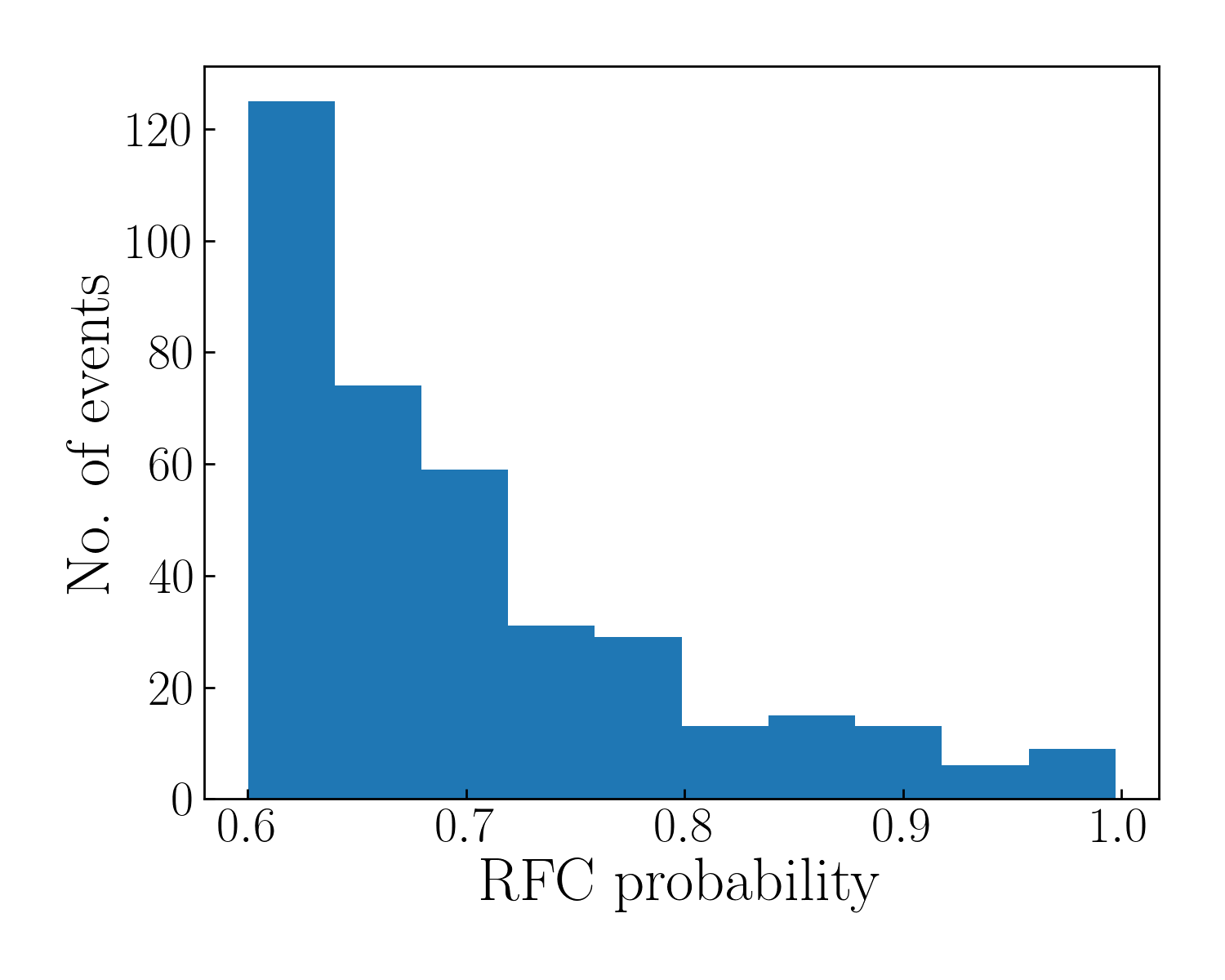}
    \caption{Histogram of probabilities for false positive candidates detected by the Random Forest Classifier with our FRB search pipeline between 10th July 2019 and 20th July 2020. Most (85\%) false positive candidates are detected with a probability of $< 0.8$), in contrast to Figure \ref{fig:RCF prob} where we find that most mock FRBs are assigned a probability higher than 0.8. See Section \ref{sec: Additional insights} for further details. }
    \label{fig:false-positives}
\end{figure}

In this section we look at additional insights resulting from the mock FRB injection, which have lead to improvements in the pipeline operation and processing efficiency.

\begin{itemize}
    \item Tuning the Random Forest Classifier threshold : the probability detection threshold used by the Random Forest Classifier has been set at 0.6 for the UTMOST FRB search program. The distribution of classification probability values for the mock FRB injections have shown that this threshold can be set higher, to $\sim$ 0.8, without significantly increasing ($<15\%$) the false-negative rate. This reduces the computational load on the FRB manager, freeing resources for other tasks. When compared with the distribution of all false positives detected by RFC in the last 12 months (Figure \ref{fig:false-positives}), this also suggests that increasing the classification threshold to 0.8 would significantly reduce the ratio of false-positive to true-positive candidates, allowing for other science projects, like  automated triggering of multi-wavelength follow-up campaigns, to work with higher efficiency.
    
    \item Effects of RFI on the FRB classifier : careful analysis of the properties of missed injections can reveal subtle biases in the machine learning based classifiers. Using a subset of mock FRBs that were injected multiple times in the live data during different observations, we discovered that the presence of an RFI event at particular frequencies in the vicinity of the injection caused those injections to be preferentially missed by the RFC in comparison to the cases when the same mock FRBs were injected in data without any RFI. Assuming that in the training of the RFC it had not been presented with enough examples of candidates in the vicinity of that particular RFI, we decided to retrain the classifier, on those missed mock FRB injections, to fix the problem.
    
    \item Finding faults in the FRB pipeline : large numbers of injections in live data can discover hidden leaks in the detection pipelines. At the end of an observation, the observing backend issues a stop signal to the data processing nodes which causes the RFC to immediately stop classifying any further candidates until a start signal for the next observation is received. Since there are some data left in the memory ring buffers waiting to be processed even after the observation has finished, any candidates in the last few ($\sim$ 3) seconds of data still remaining in the ring buffers do not get classified and are discarded at the start of next observation. As most of the observations at UTMOST are carried out for a duration of $\sim$ 360 seconds, this problem resulted in dropping of $\sim$ 1\% of candidates from every observation. Careful investigation of missed injections helped in identifying this problem which was then promptly fixed.
    
    

\end{itemize}

\section{Summary and Conclusions}
\label{sec:Conclusions}

We have developed a pipeline for injecting mock Fast Radio Bursts (FRBs) into live data streams at the Molonglo Radio Telescope, in order to characterise our FRB detection efficiencies. An injection run of 20,000 mock FRBs over the course of 6 days observing was carried out in September 2019, with a wide range of widths, dispersion measures and Signal-to-Noise ratios. The distributions are modeled as flat in linearly sampled width $w$, DM and SNR space, in the ranges $ 0 < w < 100 $ ms, $0 < $ DM $<5000$ pc cm$^{-3}$ and $6 < $ SNR $< 45$. 

A detailed analysis of the recovered FRB fractions as functions of width, DM and SNR was carried out. The study extends the analysis carried out on a smaller sample of 2,000 mock FRBs injected in March 2019, and reported in \cite{Farah2019MNRAS}. 

In the pulse width and SNR range probed in \cite{Farah2019MNRAS}, (widths $<16$ ms, SNR $>10$), we find similar results, with FRB recovery rates of order 90\%. This is a relatively small correction to make to obtain the true FRB rate on the sky from our observed FRBs. 

We find that the recovery fractions drop significantly as we probe FRBs with widths in the range $20< w < 100$ ms and SNR $< 10$. Up to 80\% of FRBs can be missed, and is a strong function of width. Recovered fractions fall rapidly below a SNR of 10. 

We have used the results from the set of 20,000 mock FRB injections to estimate the survey completeness as a function of the fluence, DM and width of FRBs.
This completeness fraction will be useful for studying FRBs through population synthesis, for example, via frbpoppy \citep{FRBpoppy_Gardenier2019}, allowing us to examine the intrinsic distribution of  FRB energetics, widths, and other properties like scattering after accounting for the bias induced by the selection function of our survey.

We draw these generalisations from our study: 
\begin{itemize}
    
    \item FRB search pipelines can have significant incompleteness and inefficiencies across their search parameter spaces. Blind mock FRB injections offer unique capability to carry out end-to-end testing of the detection pipelines, providing a means of measuring the survey completeness function.
    
    \item The FRB recovery rates can depend very strongly on particular pulse properties (such as width) and fall rapidly below the typical SNR threshold of 10. Mock FRB injections have characterised the properties of our FRB pipeline very clearly, well worth the considerable effort to implement live FRB injections into the data streams.

    \item Injections revealed rare leaks (missed FRBs) in the pipelines which are otherwise difficult to find, and offer a tool to carefully understand the inefficiencies at each step within the search pipelines.
    
    \item Injections provide an efficient means of evaluating the performance of machine-learning based FRB classification algorithms.
    
    \item Real-time mock FRB injection systems can be used to generate data-sets for training of upcoming machine-learning classifiers and fine-tuning the existing ones.
    
\end{itemize}

We encourage other observatories to undertake similar efforts towards testing their detection pipelines. Our codes are publicly available on GitHub for this purpose. 

\section*{Acknowledgment}
The authors are grateful to Evan Keane and Vincent Morello for deeply insightful discussions on Signal-to-Noise ratios. We would also like to thank Robert Main for useful comments on the paper. The Molonglo Observatory is owned and
operated by the University of Sydney, with support from the School
of Physics and the University. The UTMOST project is also supported by the Swinburne University of Technology. We acknowledge the Laureate Fellowship FL150100148. ATD is supported by
an ARC Future Fellowship grant FT150100415. Parts of this work were performed on the OzSTAR national facility at Swinburne University of Technology. OzSTAR is funded by Swinburne University of Technology and the National Collaborative Research Infrastructure Strategy (NCRIS). This research made use of numpy \citep{numpy}, pandas \citep{pandas}, matplotlib \citep{matplotlib}, jupyter \citep{jupyter}, seaborn \citep{seaborn} and PSRDADA\footnote{\href{PSRDADA}{https://sourceforge.net/projects/psrdada/}} packages.

\section*{Data availability}
The properties of all mock FRBs as injected and detected by the real-time system are compiled into a single csv file. This csv file, along with the code used to process the real-time candidates of \textsc{HEIMDALL} and the RFC, are made publicly available here: \url{https://github.com/vg2691994/mock_frb_injection_results}.
The raw injection data snippets are available for access upon request.

\bibliographystyle{mnras}
\bibliography{bibliography,python_packages}

\appendix

\section{Baseline normalisation inside \textsc{HEIMDALL}}
\label{appndix:A}

\begin{figure*}

 \begin{subfigure}[b]{.47\textwidth}
    \centering
    \includegraphics[width = \textwidth]{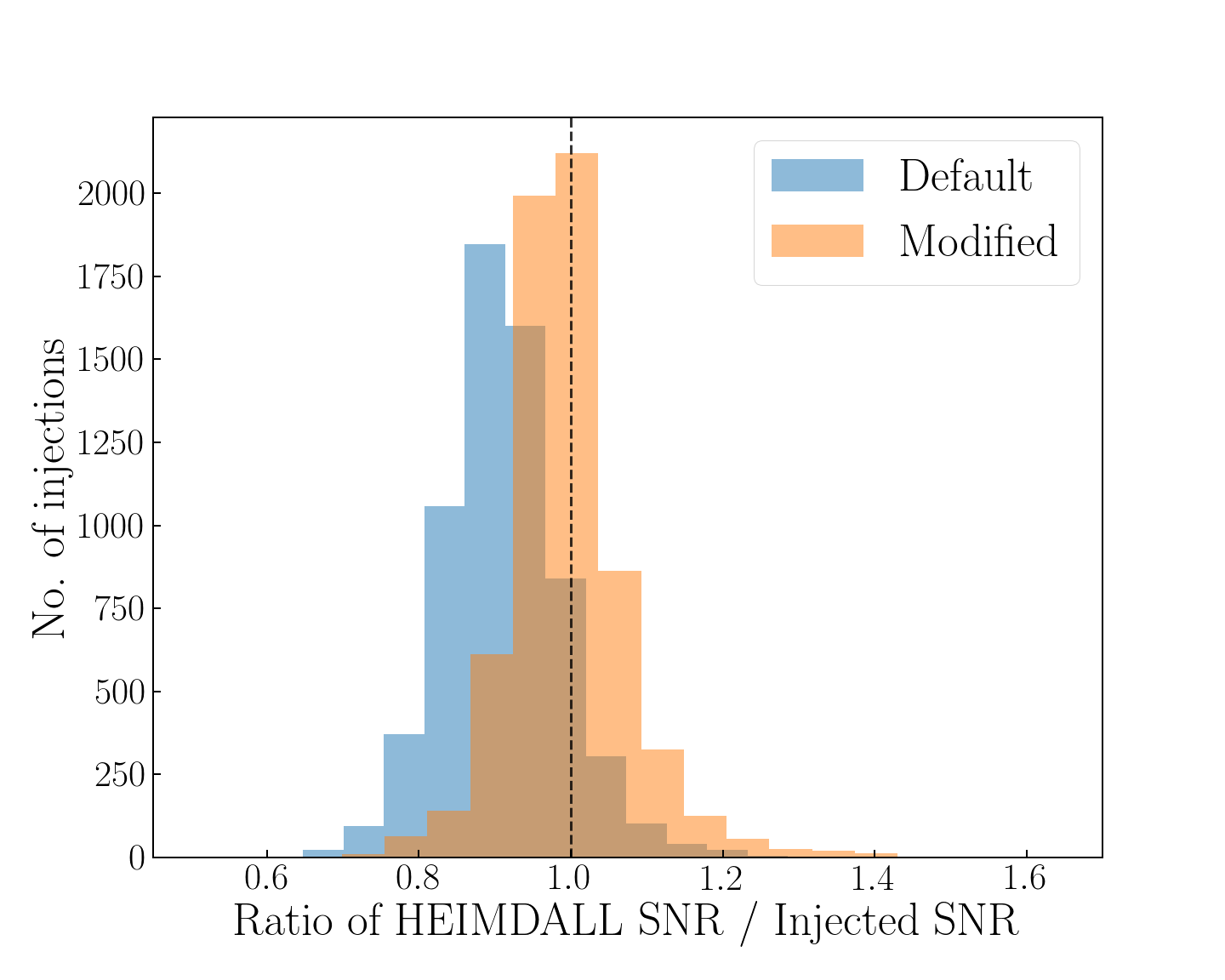}
    \caption{SNR recovery as a fraction of the injected SNR}
    \label{fig:HEIMDALL SNR recovery2}
 
 \end{subfigure}
 \begin{subfigure}[b]{.47\textwidth}
    \centering
    \includegraphics[width=\textwidth]{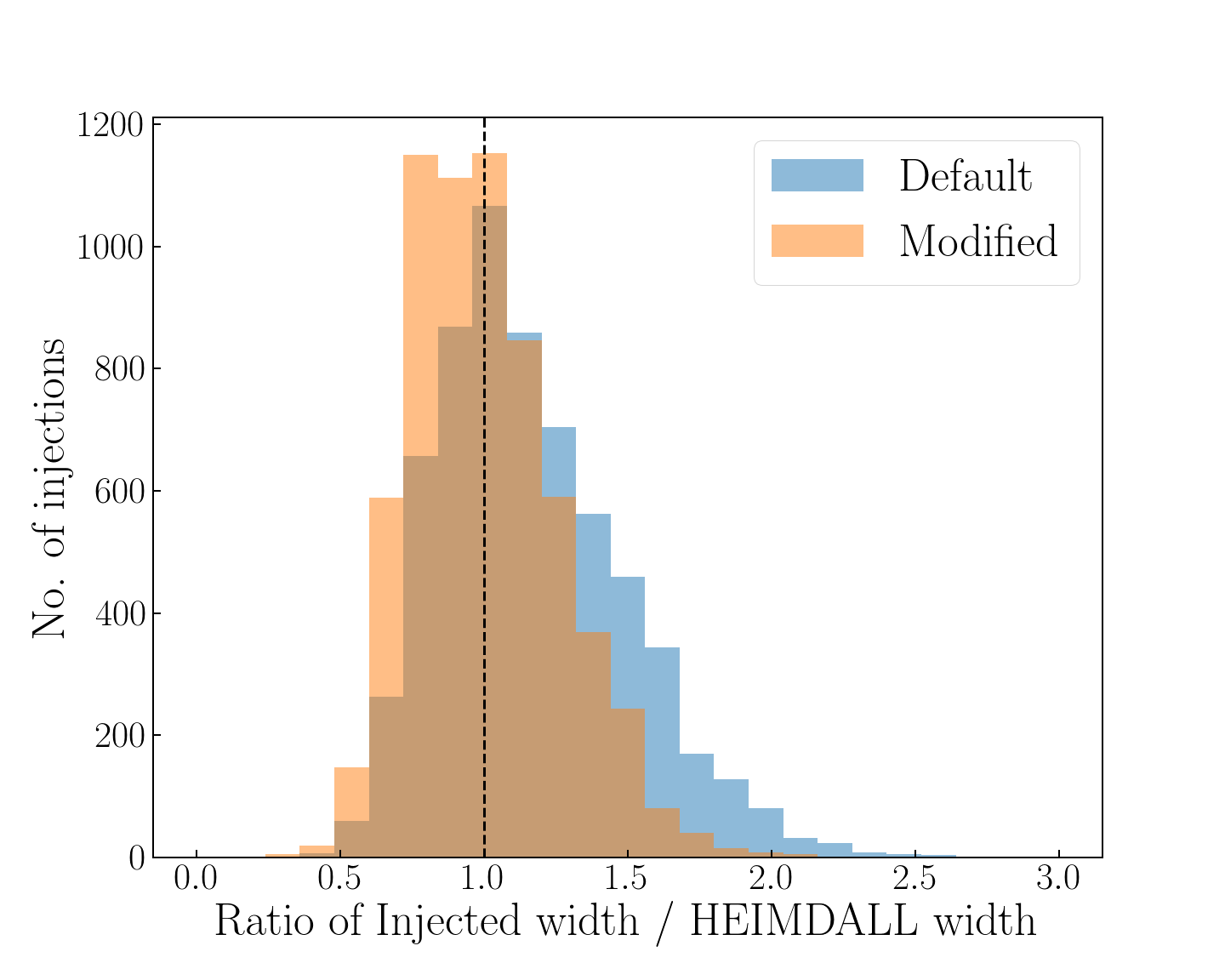}
    \caption{Width recovery as a fraction of the injected width}
    \label{fig:HEIMDALL width recovery2}
 \end{subfigure}
     
  \caption{Improvement in SNR and width recovery of mock FRBs after modification to the default rms computation method inside \textsc{HEIMDALL}. Blue histograms show the distribution of the recovered SNR and width values when a sub-sample of mock FRBs injected on simulated Gaussian noise were processed with the default rms computation method within \textsc{HEIMDALL}. Orange histograms show the distributions after processing the same injections with the modified version of \textsc{HEIMDALL}.}
  \label{fig:Heimdall fix}
\end{figure*}

First we present an overview of the SNR computation procedure of a candidate within \textsc{HEIMDALL}. The following steps are performed on each ``gulp'' of data, where a gulp is a block of frequency-time samples ingested and processed by \textsc{HEIMDALL} at each iteration (set at 720,896 samples for UTMOST):
\begin{enumerate}
    \item For a given DM trial, de-disperse the data
    \item Average across frequency to make a time series
    \item Subtract the median value of the time series baseline
    \item Measure the root-mean-square (rms) of the noise in the time series and divide it by that rms
    \item Convolve the normalised time series with box-cars of different widths corresponding to each width trial being searched
    \item Measure the rms of noise in each convolved time series and re-normalise it by dividing by the measured rms
    \item Find the sample values which exceed the detection threshold (giants) in the re-normalised time series. The value at peak of each giant represents the SNR of that candidate.
    \item Cluster giants in time and different width and DM trials. The width and DM trial at which maximum SNR is achieved are reported as the recovered width and DM values for that candidate.
\end{enumerate}

\textsc{HEIMDALL} estimates the (rms) of the noise in the time series by measuring the median absolute deviation (MAD) and scaling it by a constant scale factor $k$ ($k = 1.4826$ for Gaussian distribution). The rms estimate using the MAD is robust against bright outliers in the time series, making it the preferred method for blind searches, where it is not possible to exclude the signal from the noise.

While MAD is less susceptible than direct measurement of rms to the presence of bright events in the time series, it is still affected by outliers, including the signal and any RFI occupying a significant fraction of samples. Therefore, the presence of multiple pulses, say from a pulsar, or the presence of repetitive RFI in the data can cause the rms estimate to increase.
This bias increases as these outliers occupy an increasing fraction of the total number of samples --- which is typically the case as the boxcar convolution width is increased.
This results in an overestimation of the noise rms and a corresponding reduction in the peak values of the giants (i.e. the SNR) after re-normalisation. Convolution with wider box-cars causes a larger reduction in SNR, and hence, the highest SNR value within a cluster of giants tends to be recovered at a lower width than the true width of a signal.

To test this hypothesis, we took a sub-sample of our injections which were detected by \textsc{HEIMDALL} in the 41.94 ms width trial (shown in pink in Figure \ref{fig:HEIMDALL width recovery}) and re-injected them in a single filterbank file containing simulated Gaussian noise. We kept a gap of 9155 samples ($\approx$ 3 seconds) between every adjacent injection.
We processed this filterbank with \textsc{HEIMDALL} and measured the accuracy of the SNR and width recovery in this case. We found that the recovered SNR values were $\sim$10\% lower and the recovered width distribution was also asymmetric (Figure \ref{fig:Heimdall fix}).

We then modified the rms estimation method inside \textsc{HEIMDALL} to, instead of re-measuring the rms after convolution, compute the rms in the convolved time series by scaling the rms of the time series before convolution by a factor equal to the square-root of the width of the box-car in the convolution kernel. The same filterbank was then reprocessed with this modified version of \textsc{HEIMDALL} and analysed for its SNR and width recovery. We found dramatic improvements in the accuracy of both, SNR and width recovery, with \textsc{HEIMDALL} recovering $>99\%$ of the injected SNR and the width recovery distribution was also symmetric. These measurements are plotted in Figure \ref{fig:Heimdall fix}.

We further test this hypothesis by re-injecting the same sub-sample of mock FRBs in simulated Gaussian noise filterbanks but with different gaps between adjacent injections. After analysing \textsc{HEIMDALL}'s performance for each case, we find that the accuracy of SNR and width recovery increases with increasing gap between adjacent injections.
Therefore, we conclude that the fraction of samples lying above the median of a time series is the dominant driver of the loss in SNR. 

Since the live data at UTMOST are severely affected by RFI, with some RFI activity present up to $\sim$10\% of the observing time, we infer that the larger loss in the recovered SNR ($\sim$20\%) for live injections done at UTMOST can be explained by this sub-optimal implementation of rms estimation compounded by the harsh RFI environment at UTMOST.

We note that most current/past FRB surveys running at different radio telescopes like Parkes, GBT and STARE2, which use \textsc{HEIMDALL} as their primary detection software, would also be affected by this problem of loss in recovered SNR, but the extent to which they are affected would vary depending upon their respective RFI environments.

While it is hard to escape the increasing menace of RFI, we propose our modification to the rms estimation function inside \textsc{HEIMDALL} as a potential fix to reduce the loss in the recovered SNR caused by this problem. We note that while this modification will result in a uniform detection threshold at all widths and a more accurate recovery of candidates' SNRs/widths, it can also result in an overall increase in the number of candidates detected by \textsc{HEIMDALL}, depending upon the local RFI environment. This can increase the computation load on the classifier processing the candidates reported by \textsc{HEIMDALL}. Therefore, a trade-off between survey completeness and computational load needs to be kept in mind while making a decision to implement this change in \textsc{HEIMDALL}.







\bsp	
\label{lastpage}
\end{document}